\documentclass[british]{article}
\usepackage[left=1in,top=1in,right=1in,bottom=1in,nohead,paperwidth=8.5in, paperheight=11in]{geometry} 
\usepackage{amsmath}
\usepackage{stmaryrd}
\usepackage{natbib} 
\usepackage{float}
\usepackage{rotating}
\usepackage{multirow}
\usepackage{algorithm}
\usepackage[titletoc, title]{appendix}
\usepackage{adjustbox}
\usepackage{graphicx}
\usepackage{bbm}
\usepackage{hyperref}

\graphicspath{{./Figures/}}

\title{\bf Volatility Spillovers and Heavy Tails: \\ A Large $t$-Vector AutoRegressive Approach}

\author{Luca Barbaglia$^a$\footnote{Corresponding author: luca.barbaglia@kuleuven.be}, 			
	Christophe Croux$^a$
	and 
	Ines Wilms$^{a,c}$
	\\ 
	\textit{$^a$ \small Faculty of Economics and Business, KU Leuven, Belgium}
	\\
	\textit{$^c$ \small Cornell University, Ithaca, USA}
}

\date{}

\begin{document}
\maketitle                                                                                                              

\vspace{1cm}
{\bf Abstract.}
Volatility is a key measure of risk in financial analysis. 
The high volatility of one financial asset today could affect the volatility of another asset tomorrow. 
These lagged effects among volatilities - which we call volatility spillovers - are studied using the Vector AutoRegressive (VAR) model.
We account for the possible fat-tailed distribution of the VAR model errors  using a VAR model with errors following a \textit{multivariate Student $t$}-distribution with unknown degrees of freedom.
Moreover, we study volatility spillovers among a \textit{large} number of assets. To this end, we use \textit{penalized} estimation of the VAR model with $t$-distributed errors.
We study volatility spillovers among  energy, biofuel and agricultural commodities
and reveal bidirectional volatility spillovers between energy and biofuel, and between energy and agricultural commodities.


\bigskip



\newpage

\section{Introduction \label{VOL_intro}}
Volatilities of financial asset returns are closely followed by financial market analysts and investors. 
As volatility is a risk measure, analysts working in portfolio management are interested in better understanding the volatility of the financial assets in their portfolio in order to minimize the exposure to future potential losses.
Knowing whether the high volatility of one financial asset today leads to high volatility of another  asset tomorrow can be key information to reduce this risk exposure. 
We focus on these lagged effects among volatilities and we refer to them as volatility spillovers.

We investigate the existence of volatility spillovers among  energy, biofuel and agricultural commodities (see \citealp{Serra13}, and references therein). 
Commodity markets are of great interest for financial analysts and investors, who uses them for risk hedging purposes or as alternative investment areas.
In recent years, the high volume of financial transactions in commodity markets has lead to a detachment from simple supply-demand dynamics. 
This makes it harder for analysts to describe swings in  volatilities simply recurring to the economic fundamentals. 
In addition, commodity markets are volatile by their very nature: a policy change, a natural disaster or a technological breakthrough might cause a unexpected period of high volatility.

Traditionally, high volatility in the energy markets would affect unidirectionally the volatility of agricultural products via energy intensive agricultural inputs (e.g., fuel and fertilizers).
The emergence of biofuel production has changed the link between energy and agricultural commodities as certain crops represent simultaneously food, feedstock and fuel sources.
A joint modeling of the volatility spillovers among energy, biofuel and agricultural commodities is therefore relevant, as it might improve  traditional risk management tools and lead to a better assessment of biofuel support policies \citep{Serra11}.

We follow \cite{Diebold15} and analyze volatility spillovers using the Vector AutoRegressive (VAR) model. 
First, we obtain a measure of volatility (e.g. see \citealp{Martens07} or \citealp{McAleer08}).
Then, we consider a VAR model containing the logarithmic transformed volatilities as time series to estimate the spillovers.
This procedure has been used in many empirical applications, but two main concerns remain:
(i) the standard estimation procedure of the VAR does not account for fat-tailed errors and thus for the frequent occurrence of extreme observations in the volatility series (e.g. \citealp{Callot17}) and,  
(ii) the number of time series in the VAR model is limited since the number of parameters to be estimated increases quadratically with the number of time series included (\citealp{Diebold15}, p183).

Several recent papers propose methods to estimate large VAR models - that is models containing a large number of time series relative to the time series length -  (e.g. \citealp{Basu15, Davis16, Gelper16, Barigozzi17, Demirer17}).
The  extension of large VAR models to fat-tailed errors has, however, not been addressed yet.
It has been argued that log volatilities are approximately normal (\citealp{Diebold15}, p18), which is supported by empirical evidence (e.g. \citealp{Andersen01ST}) and  asymptotic theory \citep{BarndorffNielsen02}. 
Nevertheless, subsequent studies emphasize that, in practice, this might not hold for all volatility measures. 
Considering realized variances, \cite{Corsi08} and \cite{Hassler16} highlight that the log volatility of the S\&P 500 index based on 5min intra-day returns exhibits deviations from the Gaussian distribution.
Considering realized ranges, \cite{Christensen09} prove that the log volatility follows a mixed multivariate normal, and \cite{Caporin15} show that the log volatility has heavier tails than the normal.
Hence, it is appropriate to specify a more general distribution for the errors of the VAR model.

We analyze volatility spillovers using a large $t$-VAR estimation approach. 
To this end, we rely on the work of \cite{Diebold15}, but we (i) use a  VAR model with \textit{$t$-distributed errors}.
We \textit{estimate the degrees of freedom} of the $t$-distribution, in contrast to prior studies  who take them to be fixed (e.g. \citealp{Franses98}; \citealp{Finegold11}). As such, we determine the degree of fat-tailedness of the VAR residuals in a data-driven manner.
(ii) We consider \textit{large} $t$-VAR models containing a large number of time series relative to the time series length.
To ensure that the estimation of this large VAR model is feasible, we use a penalized estimator of the VAR model in the spirit of the Lasso \citep{Tibshirani96} and call it the \textit{t-Lasso}. The $t$-Lasso attains higher estimation and forecast accuracy in presence of fat-tailed errors than other standard estimators, as confirmed by our simulation study.

The results from the commodity analysis reveal the fat-tailedness of the VAR innovations, since the estimated degrees of freedom of the $t$-distribution are very low. 
It turns out that the  $t$-Lasso estimator, that accounts for this fat-tailedness,  improves forecast accuracy.
In our commodity analysis, we find volatility spillovers between energy and biofuels. 
Among energy commodities, most of the volatility spillovers involve gasoline, which is often blended with ethanol for consumption.
Moreover, we observe volatility spillovers between energy and agriculture, regardless of the fact whether those crops can be used for biofuel production or not.
We also find bidirectional volatility spillovers between biofuels and those agricultural commodities that can be used as inputs in biofuel production.
These volatility spillovers are not observed in times of low energy prices, when investing in biofuels is less profitable.

The remainder of this article is structured as follows.
Section \ref{VOL_model} reviews the VAR model and introduces the corresponding $t$-Lasso estimator.
Section \ref{VOL_algorithm} outlines the algorithm.
The simulation study  in Section \ref{VOL_simulations} shows the good performance of the proposed estimator.
Section \ref{VOL_Network_analysis} presents the data, the definition of volatility spillovers and the 
network analysis tool to visualize them.
Section \ref{VOL_DataResults}  presents the results of the volatility spillover analysis.
Finally, Section \ref{VOL_Discussion} concludes.

\section{Models and estimators \label{VOL_model}}

Let $\boldsymbol{y}_{t}$ be a $J$-dimensional vector of log volatilities for $1 \leq t \leq T$, with $T$ the time series length. 
We take the logarithmic transformation of the volatilities, which ensures the positivity of the volatility forecasts (e.g. \citealp{Callot17}). 
We consider a stationary VAR of order $P$ for the log volatilities
\begin{equation}
\boldsymbol{y}_t = \sum_{p=1}^{P} \boldsymbol{B}_{p} \boldsymbol{y}_{t-p} + \boldsymbol{e}_{t}\label{VOL_VARp}, 
\end{equation}
where $\boldsymbol{B}_{p}$ is the $J \times J$ matrix of autoregressive coefficients for lags $1 \leq p \leq P$, and $\boldsymbol{e}_t$ is a $J$-dimensional vector of error terms with zero mean and covariance matrix $\boldsymbol{\Sigma}$.
Without loss of generality, we assume that all log volatility time series are mean centered such that no intercept is included.

For ease of notation, we rewrite model \eqref{VOL_VARp} in matrix form
\begin{equation}
\boldsymbol{Y}= \boldsymbol{X}\boldsymbol{B} + \boldsymbol{E}, \label{VOL_VARp_matrix} \nonumber
\end{equation}
where the $N \times J$ matrix $\boldsymbol{Y}=[\boldsymbol{y}_{P+1}', \ldots, \boldsymbol{y}_{T}']'$, with $N=T-P$.
Let $\boldsymbol{X}=[\boldsymbol{X}_{1}, \dots, \boldsymbol{X}_{P}]$ be the $N \times JP$ matrix of lagged time series, where $\boldsymbol{X}_{p}=[\boldsymbol{y}_{P+1-p}', \ldots, \boldsymbol{y}_{T-p}']'$ is a $N \times J$ matrix for $1 \leq p \leq P$. 
Finally, the $JP \times J$ matrix of autoregressive coefficients is $\boldsymbol{B}=[\boldsymbol{B}_{1}', \ldots, \boldsymbol{B}_{P}']'$ and $\boldsymbol{E}$ is the $N \times J$ error matrix.

In the next subsections we first review the penalized estimator for the Gaussian VAR model, and then we
introduce the penalized estimator for the VAR model with Student $t$-errors.
To this end, we build on the $t$-Lasso estimator for sparse graphical models proposed by \cite{Finegold11}.
We extend the approach of \cite{Finegold11} by 
(i) allowing for an unknown degrees of freedom of the $t$-distribution 
(ii) by adding a penalty on the inverse covariance matrix, ensuring existence of the estimator in any dimension (iii) extending their static framework for interdependencies between responses to the VAR model.

\subsection{Gaussian innovations\label{VOL_Gaussian_model}}
Assume that the error terms $\boldsymbol{e}_t$ follow a multivariate normal distribution $N(\boldsymbol{0}, \boldsymbol{\Sigma})$.
Consider the following penalized least squares estimator of $\boldsymbol{B}$ 
\begin{equation}
\widehat{\boldsymbol{B}}= \underset{\boldsymbol{B}}{\operatorname{argmin}} \ \dfrac{1}{2N} 
\operatorname{tr}\left[(\boldsymbol{Y} - \boldsymbol{X}\boldsymbol{B})(\boldsymbol{Y} - \boldsymbol{X}\boldsymbol{B})'\right]
+ \lambda \sum_{i,j=1}^{J} \sum_{p=1}^{P} |B_{p,ij}|, \label{VOL_GrLasso}
\end{equation}
where $\operatorname{tr}(\cdot)$ is the trace operator, $\lambda>0$ is a regularization parameter associated to the $l_{1}$-penalty (\citealp{Hastie15}, p8) on the $ij^{th}$ entry of $\boldsymbol{B}_{p}$ denoted by $[\boldsymbol{B}_{p}]_{ij}=B_{p,ij}$. 
This penalty ensures that the estimation of $\boldsymbol{B}$ is feasible  even if the number of parameters exceeds the time series length. 
Moreover, it sets some elements of $\widehat{\boldsymbol{B}}$ exactly equal to zero.
The larger the regularization parameter $\lambda$, the sparser  $\widehat{\boldsymbol{B}}$, that is the more of its elements are exactly zero.

In line with \cite{Rothman10}, we account for correlated errors by including the estimation of the inverse error covariance matrix $\boldsymbol{\Omega}=\boldsymbol{\Sigma}^{-1}$. To this end, we turn to the maximum likelihood framework and jointly estimate $\boldsymbol{B}$ and $\boldsymbol{\Omega}$ by minimizing the negative log likelihood: 
\begin{equation}
(\widehat{\boldsymbol{B}}, \widehat{\boldsymbol{\Omega}})= \underset{\boldsymbol{B},\boldsymbol{\Omega}}{\operatorname{argmin}} \ \dfrac{1}{2N} 
\operatorname{tr}\left[(\boldsymbol{Y} - \boldsymbol{X}\boldsymbol{B})\boldsymbol{\Omega}(\boldsymbol{Y} - \boldsymbol{X}\boldsymbol{B})'\right]
-\dfrac{1}{2}\log|\boldsymbol{\Omega}|
+ \lambda \sum_{i,j=1}^{J} \sum_{p=1}^{P} |B_{p,ij}|
+ \gamma \sum_{i \neq j}^{J}|\omega_{ij}|, \label{VOL_JGrLasso}
\end{equation}
where $\omega_{ij}$ is the $ij^{th}$ element of $\boldsymbol{\Omega}$, and $\gamma>0$ is the regularization parameter associated to the $l_{1}$-penalty on the off-diagonal elements of the inverse error covariance matrix \citep{Friedman08}. This $l_{1}$-penalty ensures that the estimation of $\boldsymbol{\Omega}$ is feasible even if the number of parameters exceeds the time series length. 
Furthermore, it sets some elements of $\widehat{\boldsymbol{\Omega}}$ to zero. 
The larger the regularization parameter $\gamma$, the sparser  $\widehat{\boldsymbol{\Omega}}$.	
We refer to the estimator in \eqref{VOL_JGrLasso} as the \textit{Gaussian Lasso}.

\subsection{Student \textit{t}-innovations\label{VOL_student_t_subsection}}
We depart from the normality assumption and assume that the error terms are distributed according to a multivariate $t$-distribution $t_{\nu}(\boldsymbol{0}, \boldsymbol{\Psi})$, where $\nu>0$ are the degrees of freedom and $\boldsymbol{\Psi}$ is the scale matrix, with associated variance-covariance matrix $\boldsymbol{\Sigma}=\boldsymbol{\Psi}\nu/(\nu-2)$ if $\nu>2$.  
The associated density function is given by
\begin{equation}
\dfrac{\Gamma((\nu + J)/2)|\boldsymbol{\Omega}|^{1/2}}
{(\pi\nu)^{J/2}\Gamma(\nu/2)[1+\boldsymbol{e}_{t}' \ \boldsymbol{\Omega} \ \boldsymbol{e}_{t}/\nu]
	^{(\nu+J)/2}}\label{VOL_tdist}, 
\end{equation}
with $\boldsymbol{\Omega}=\boldsymbol{\Psi}^{-1}$ (\citealp{Kotz04}, p1). 
Recall that $J$ is the dimension of the time series.

The Student $t$-distributed random vector $\boldsymbol{e}_{t}$ can be written as a Gaussian scale-mixture of $\boldsymbol{\phi}_{t}$ and $\tau_{t}$ (e.g. \citealp{Kotz04}, p2).
Here $\boldsymbol{\phi}_{t}$ is a $J$-dimensional random vector distributed as a multivariate normal $N(\boldsymbol{0},\boldsymbol{\Psi})$, independent of the random variable $\tau_{t}$ distributed as a Gamma $\Gamma(\nu/2, \nu/2)$. 
Then $\boldsymbol{e}_{t}=\boldsymbol{\phi}_{t}/\sqrt{\tau_{t}}$ follows a $t$-distribution $t_{\nu}(\boldsymbol{0}, \boldsymbol{\Psi})$. 
One can show that $\boldsymbol{e}_{t}$ has a normal conditional distribution
\begin{equation}
\boldsymbol{e}_{t}|  \tau_{t}  \sim N \left(\boldsymbol{0}, \  \dfrac{\boldsymbol{\Psi}}{\tau_{t}} \right) \label{VOL_conditionalN}
\end{equation}
and
\begin{equation}
\tau_{t} | \boldsymbol{e}_{t} \sim \Gamma\left( \dfrac{\nu+J}{2},  \ \dfrac{\nu+\boldsymbol{e}_{t}' \ \boldsymbol{\Omega} \ \boldsymbol{e}_{t}}{2} \right).\label{VOL_ExpectedTau}
\end{equation}
The arguments of the Gamma distribution are respectively the shape and scale parameters.
By the properties of the Gamma distribution
\begin{equation}
E\left[\tau_{t} | \boldsymbol{e}_{t} \right]= \dfrac{\nu+J}{\nu+\boldsymbol{e}_{t}' \ \boldsymbol{\Omega} \ \boldsymbol{e}_{t}}.\label{VOL_tau_expectation}
\end{equation}
The joint estimator of $\boldsymbol{B}$ and $\boldsymbol{\Omega}$ 
is defined as in \eqref{VOL_JGrLasso}, replacing the Gaussian by the $t$-density given in \eqref{VOL_tdist}, and keeping the penalty terms.
We call this estimator the \textit{$t$-Lasso}.

\section{Algorithm \label{VOL_algorithm}}
We first assume that the degrees of freedom $\nu$ is known and discuss the EM (Expectation Maximization) algorithm  to obtain the $t$-Lasso estimator, following \cite{Finegold11} for graphical models.
However, in practice $\nu$ is not known and needs to be estimated. 
To this end, we use the ECM (Expectation Conditional Maximization) algorithm to include the estimation of $\nu$ \citep{Liu95}.
The code of the algorithm is made available on the author's website. 

\subsection{EM algorithm with known $\nu$}
We treat $\tau_t$ as a hidden variable and want to estimate $\boldsymbol{B}$ and  $\boldsymbol{\Omega}$. 
In the E-step, we compute the conditional expectation of the hidden variables according to \eqref{VOL_tau_expectation}.
These expected values are put on the diagonal of an $N \times N$ diagonal matrix $\boldsymbol{\tau}$. 
In the M-step, we solve the optimization problem 
\begin{equation}
(\widehat{\boldsymbol{B}}, \widehat{\boldsymbol{\Omega}} \ | \boldsymbol{\tau})= 
\underset{\boldsymbol{B}, \boldsymbol{\Omega}}{\operatorname{argmin}} \ \dfrac{1}{2N} 
\operatorname{tr}\left[\boldsymbol{\tau}(\boldsymbol{Y} - \boldsymbol{X}\boldsymbol{B}) \ \boldsymbol{\Omega} \ (\boldsymbol{Y} - \boldsymbol{X}\boldsymbol{B})'\right]
-\dfrac{1}{2}\log|\boldsymbol{\Omega}|
+ \lambda \sum_{i,j=1}^{J} \sum_{p=1}^{P} |B_{p,ij}|
+ \gamma \sum_{i \neq j}^{J}|\omega_{ij}|, \label{VOL_JGrLasso_weighted}
\end{equation}
where the conditional normality \eqref{VOL_conditionalN} of the error terms, given the hidden variables,  is used. 
This M-step corresponds exactly to solving  problem \eqref{VOL_JGrLasso} for weighted observations. 
Algorithm 
1
gives the details of the EM algorithm.

\begin{algorithm}[t]
	\caption{\hspace{-0.15cm}\textbf{1} \ $t$-Lasso with known $\nu$: Expectation Maximization (EM)}\label{VOL_EMalgorithm}
	\begin{description}
		\item [{Input}] $\textbf{Y}$, $\textbf{X}$, degrees of freedom ${\nu}$ and desired accuracy $\varepsilon$. 
		\item [{Initialization}] 
		 $\widehat{\boldsymbol{\Omega}}^{(0)}=\textbf{I}_{J}$, $\widehat{\boldsymbol{B}}_{p}^{(0)}=\boldsymbol{I}_{J}$ for $1 \leq p \leq P$,
		and  $\boldsymbol{e}^{(0)}_t$ the $t^{th}$ row of 
		$\boldsymbol{Y}-\boldsymbol{X}\boldsymbol{\widehat{B}}^{(0)}$.
		\item[{Iteration}] Iterate the following steps for $m=0,1,2,\ldots$:
		\begin{description}
			\item [{E-step}] Compute for $t=1, \ldots, N$ the weights
			\[
			\widehat{\tau_t}^{(m+1)}=
			\dfrac{\nu+J}{\nu+ 
				(\boldsymbol{e}^{(m)}_t)'
				\boldsymbol{\widehat{\Omega}}^{(m)} 
				\boldsymbol{e}^{(m)}_{t}
				}.
			\] 	
			\item [{M-step}] Compute $\widehat{\boldsymbol{B}}^{(m+1)}$ and $\widehat{\boldsymbol{\Omega}}^{(m+1)}$  using Algorithm 
			A
			in Appendix
			A
			with inputs $\textbf{Y}=\widehat{\boldsymbol{\tau}}^{*(m+1)}\boldsymbol{Y}$, and $\textbf{X}=\widehat{\boldsymbol{\tau}}^{*(m+1)}\boldsymbol{X}$.
			Here, $\widehat{\boldsymbol{\tau}}^{*(m+1)}$ is the diagonal matrix having the square root of $\widehat{\tau}_t^{(m+1)}$ on its diagonal, for $t=1, \ldots, N$.
			\item[] \hspace{-0.25cm} Let $\boldsymbol{e}^{(m+1)}_t$ be the $t^{th}$ row of 
			$\boldsymbol{Y}-\boldsymbol{X}\boldsymbol{\widehat{B}}^{(m+1)}$.
		\end{description}
		\item [{Convergence}] Iterate until the relative change in the value of the objective function in \eqref{VOL_JGrLasso_weighted} in two successive iterations is smaller than $\varepsilon$.
		\item[Output] $\widehat{\boldsymbol{B}}=\widehat{\boldsymbol{B}}^{(m+1)}$ and  $\widehat{\boldsymbol{\Omega}}=\widehat{\boldsymbol{\Omega}}^{(m+1)}$.
	\end{description}
\end{algorithm}

\subsection{ECM algorithm with unknown $\nu$}
For unknown degrees of freedom $\nu$, the M-step is replaced by two constrained maximizations (CM) steps, where first $(\boldsymbol{B}, \boldsymbol{\Omega})$ and then $\nu$ are estimated. 
An E-step is introduced before each CM-step, such that the weights are estimated twice in each iteration. This results is a multi-cycle version of the EM algorithm.  
Conditional on the latent variable $\tau_t$ for $t=1, \ldots, N$, the log-likelihood of the parameters $\boldsymbol{B}$, $\boldsymbol{\Omega}$ and $\nu$, ignoring constants, is:
\begin{equation}
  L(\boldsymbol{B}, \boldsymbol{\Omega}, \nu|\boldsymbol{\tau})=
  L_N(\boldsymbol{B}, \boldsymbol{\Omega}|\boldsymbol{\tau})+
  L_G(\nu|\boldsymbol{\tau}),\label{VOL_loglik_ECM}
\end{equation} 
where $ L_N(\boldsymbol{B}, \boldsymbol{\Omega}|\boldsymbol{\tau})$ is the objective function in \eqref{VOL_JGrLasso_weighted}, and 
\begin{equation}
	L_G(\nu|\boldsymbol{\tau})=
	-N \log \Gamma (\dfrac{\nu}{2})
	+ \dfrac{N\nu}{2} \log (\dfrac{\nu}{2}) 
	+\dfrac{\nu}{2}\sum_{t=1}^{N}(\log(\tau_t)-\tau_t). \label{VOL_Lgamma}
\end{equation}
Given $\boldsymbol{B}$ and $\boldsymbol{\Omega}$, and assuming that the latent variable $\tau_{t}$ for $t=1, \ldots, N$ is observed,
we optimize \eqref{VOL_Lgamma} and obtain $\nu$ as the solution of
\begin{equation}
	-\varphi(\frac{\nu}{2})+\log(\frac{\nu}{2})+
	\frac{1}{N}\sum_{t=1}^{N}\left(\log(\tau_{t})-\tau_{t}\right)+1=0, \label{VOL_ECM_nu_basic}
\end{equation}
where $\varphi(\cdot)$ is the derivative of the log Gamma function. 
However, in practice $\tau_t$ is not observed. Thus, we replace the term  $\sum_{t=1}^{N}\left(\log(\tau_{t})-\tau_{t}\right)$ with its expectation \citep{Liu95}
\begin{equation}
	E[\sum_{t=1}^{N}\left(\log(\tau_{t})-\tau_{t}\right)|\boldsymbol{B},\boldsymbol{\Omega},\nu]
	=	\frac{1}{N}\sum_{t=1}^{N}\left(\log(\widehat{\tau}_t)-\widehat{\tau}_t\right)
	+\frac{1}{N}\left[\varphi\left(\frac{\nu+J}{2}\right)-\log\left(\frac{\nu+J}{2}\right)\right],\label{VOL_ExpectationSUM_ECM}
\end{equation}
where we use \eqref{VOL_ExpectedTau} and where $\widehat{\tau}_t$ is  computed as in \eqref{VOL_tau_expectation}.
We substitute \eqref{VOL_ExpectationSUM_ECM} in \eqref{VOL_ECM_nu_basic}, and estimate $\nu$ as the solution of
\begin{equation}
	-\varphi(\frac{\nu}{2})+\log(\frac{\nu}{2})
	+\frac{1}{N}\sum_{t=1}^{N}\left(\log(\widehat{\tau}_t)-\widehat{\tau}_t\right)+1 
	+\frac{1}{N}\left[\varphi\left(\frac{\nu+J}{2}\right)-\log\left(\frac{\nu+J}{2}\right)\right] = 0.
	\label{VOL_solution_df}
\end{equation}
Algorithm 
2
presents the complete ECM algorithm.

\begin{algorithm}[t]
	\caption{\hspace{-0.15cm}\textbf{2} \ $t$-Lasso with unknown $\nu$: Expectation Conditional Maximization (ECM)}\label{VOL_MCECMalgorithm}
	\begin{description}
		\item [{Input}] $\textbf{Y}$, $\textbf{X}$, and desired accuracy $\varepsilon$. 
		\item [{Initialization}] 
		$\widehat{\boldsymbol{\Omega}}^{(0)}=\textbf{I}_{J}$, $\widehat{\boldsymbol{B}}_{p}^{(0)}=\boldsymbol{I}_{J}$ for $1 \leq p \leq P$,
		$\boldsymbol{e}^{(0)}_t$ the $t^{th}$ row of 
		$\boldsymbol{Y}-\boldsymbol{X}\boldsymbol{\widehat{B}}^{(0)}$, 
		and $\widehat{\nu}^{(0)}=1000$.
		\item[{Iteration}] Iterate the following steps for $m=0,1,2,\ldots$:
		\begin{description}
			\item [{E-step 1}] Compute for $t=1, \ldots, N$ the weights:
			\[
			\widehat{\tau_t}^{(m+\frac{1}{2})}=
			\dfrac{\widehat{\nu}^{(m)}+J}{\widehat{\nu}^{(m)}+ 
				(\boldsymbol{e}^{(m)}_t)'
				\boldsymbol{\widehat{\Omega}}^{(m)}
				\boldsymbol{e}^{(m)}_{t}
			}.
			\]
			\item [{CM-step 1}]  Compute $\widehat{\boldsymbol{B}}^{(m+1)}$ and $\widehat{\boldsymbol{\Omega}}^{(m+1)}$  using 
			Algorithm
			A 
			with inputs $\textbf{Y}=\widehat{\boldsymbol{\tau}}^{*(m+\frac{1}{2})}\boldsymbol{Y}$, and $\textbf{X}=\widehat{\boldsymbol{\tau}}^{*(m+\frac{1}{2})}\boldsymbol{X}$.
			Here, $\widehat{\boldsymbol{\tau}}^{*(m+\frac{1}{2})}$ is the diagonal matrix having the square root of $\widehat{\tau}_t^{(m+\frac{1}{2})}$ on its diagonal, for $t=1, \ldots, N$.
			\\
			Let $\boldsymbol{e}^{(m+1)}_t$ be the $t^{th}$ row of 
			$\boldsymbol{Y}-\boldsymbol{X}\boldsymbol{\widehat{B}}^{(m+1)}$.
			\item [{E-step 2}] Compute for $t=1, \ldots, N$ the weights:
			\[
			\widehat{\tau_t}^{(m+1)}=
			\dfrac{\widehat{\nu}^{(m)}+J}{\widehat{\nu}^{(m)}+ 
				(\boldsymbol{e}^{(m+1)}_t)'
				\boldsymbol{\widehat{\Omega}}^{(m+1)}
				\boldsymbol{e}^{(m+1)}_{t}
			}.	
			\]
			\item [{CM-step 2}] Compute $\widehat{\nu}^{(m+1)}$ with a one-dimensional search minimizing \eqref{VOL_solution_df}.
				
		\end{description}
		\item [{Convergence}] Iterate until the relative change in the value of the objective function in \eqref{VOL_JGrLasso_weighted} in two successive iterations is smaller than $\varepsilon$.
		\item[Output] $\widehat{\boldsymbol{B}}=\widehat{\boldsymbol{B}}^{(m+1)}$,  $\widehat{\boldsymbol{\Omega}}=\widehat{\boldsymbol{\Omega}}^{(m+1)}$ and
		$\widehat{\nu}=\widehat{\nu}^{(m+1)}$.
	\end{description}
\end{algorithm}	

\section{Simulations\label{VOL_simulations}}
We analyze the performance of the \textit{$t$-Lasso}, computed as outlined in Section \ref{VOL_algorithm}. 
We use both the $t$-Lasso with fixed degrees of freedom (at the true value)  and with estimated degrees of freedom. Their performance is compared to two alternative estimators:
the \textit{Least Squares} (LS), and 
the \textit{Gaussian Lasso}, i.e. the solution of equation \eqref{VOL_JGrLasso}.
The LS is the standard non-penalized estimator,
the Gaussian Lasso is the benchmark for Gaussian models.  

\paragraph{Data generating process.}
We simulate from a VAR of order $P=2$ with $J=10$ time series. The dimensions of the VAR are in line with the ones in the application to be discussed in Section \ref{VOL_Network_analysis}. The data generating process is:
\[
\boldsymbol{y}_t =  \boldsymbol{B}_{1} \boldsymbol{y}_{t-1} + \boldsymbol{B}_{2} \boldsymbol{y}_{t-2} + \boldsymbol{e}_{t},
\]
for $P+1 \leq t \leq T=100$. 
The autoregressive coefficient matrices are highly sparse in a structured manner: $B_{1}$ and $B_{2}$ have the same sparsity structure with non-zero elements ($0.4$ for $B_{1}$ and $0.2$ for $B_{2}$) on the main diagonal and on the first row. 
The error terms $\boldsymbol{e}_{t}$ follow a multivariate Student $t_{\nu}(\boldsymbol{0}, \boldsymbol{\Psi})$, with $\nu \in \{1, 2, 3, 5, 10, \infty\}$. 
In the special case $\nu=1$, the distribution is a multivariate Cauchy distribution, whereas for $\nu \rightarrow \infty$ the distribution is a multivariate normal.
The $ij^{th}$ entry of $\boldsymbol{\Psi}$ is $\psi_{i,j}=0.1^{|i-j|}$, such that inverse error covariance matrix $\boldsymbol{\Omega}$ is a band matrix.
We take $S=1000$ simulations runs.

\paragraph{Performance measures.}
The different estimators are compared in terms of their estimation accuracy. 
To evaluate the estimation accuracy, we use the Mean Absolute Estimator Error 
\[
\mathrm{MAEE}(\boldsymbol{B},\widehat{\boldsymbol{B}})=
\dfrac{1}{S}\dfrac{1}{PJ^{2}}\sum_{S=1}^{S}\sum_{i,j=1}^{J}\sum_{p=1}^{P}
|\widehat{B}_{p,ij}^{(s)}-B_{p,ij}^{(s)}|,
\]
where $\widehat{B}_{p,ij}^{(s)}$ is the $ij^{th}$ entry of the estimate $\widehat{\boldsymbol{B}}_{p}^{(s)}$ in simulation run $s$.


\paragraph{Simulation results.}
Table \ref{VOL_performance_simulations} reports the MAEEs for the four estimators and the different values of the true degrees of freedom $\nu$. 
The $t$-Lasso with $\nu$ fixed at the true value always achieves the best MAEE. 
It is very closely followed by the $t$-Lasso with  $\nu$ estimated. There is no considerable loss in estimation accuracy when estimating $\nu$. 
Both $t$-Lasso estimators perform significantly better than the Gaussian Lasso (the difference in estimation accuracy is tested with a paired $t$-test, all $p$-values $<0.01$). 
For instance,  the improvement in estimation accuracy is of $53\%$ and of $9\%$ for $\nu=1$ and $\nu=2$ respectively.
The margin by which the $t$-Lasso estimators outperform the Gaussian Lasso decreases for larger degrees of freedom.
In particular, for $\nu = \infty$ there is no significant difference between the $t$-Lasso estimators and the Gaussian Lasso. Indeed, as $\nu \rightarrow \infty$, a multivariate $t$-distribution can be viewed as the approximation of a multivariate normal.
The $t$-Lasso estimators also significantly outperform the LS for all values of $\nu$: they improve estimation accuracy by, for instance, 18\% for $\nu=3$.
The LS suffers from the large number of parameters to estimate, given the length of the time series.


\begin{table}
	\caption{Mean Absolute Estimation Error for the four estimators and different degrees of freedom $\nu$.\label{VOL_performance_simulations}}
	\medskip
	\centering
	\begin{tabular}{c|cccc}
		\hline
		$\nu$ & \small{LS} & \small{Gaussian}  & \multicolumn{2}{c}{\small{$t$-Lasso}} \\ 
		&  &  \small{Lasso} & \small{$\nu$ fixed }& \small{$\nu$ estimated} \\ 
		\hline
		1 & 0.638 & 0.187 & 0.088 & 0.089 \\ 
		2 & 0.135 & 0.096 & 0.088 & 0.088 \\ 
		3 & 0.107 & 0.090 & 0.088 & 0.089 \\ 
		5 & 0.098 & 0.090 & 0.088 & 0.089  \\ 
		10 & 0.095 & 0.091 & 0.089 & 0.090 \\ 
		$\infty$ & 0.093 & 0.091 & 0.091 & 0.091 \\ 
		\hline
	\end{tabular}    
	

\end{table}
 
\begin{figure}
	\centering
	\includegraphics[scale=0.4]{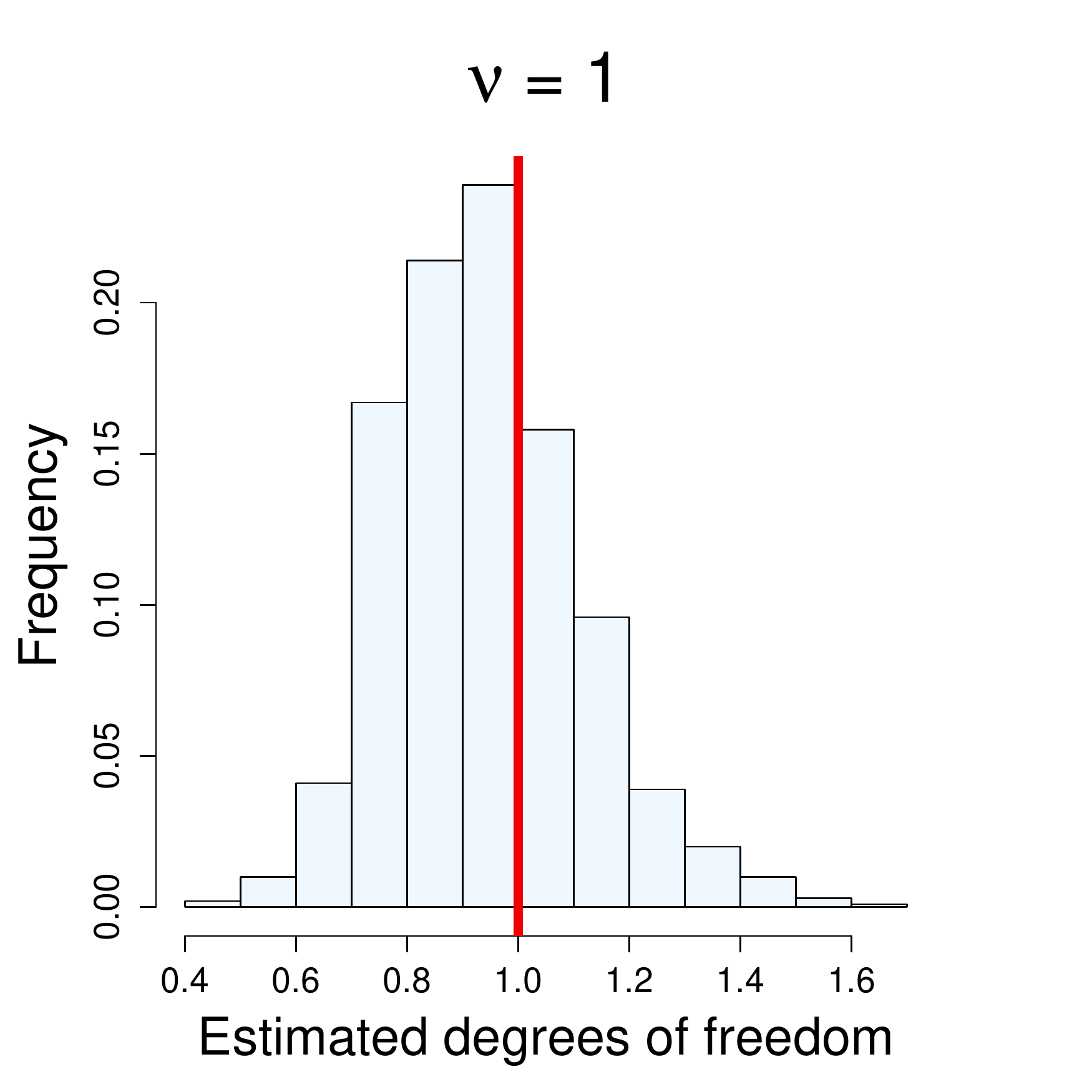} 
	\includegraphics[scale=0.4]{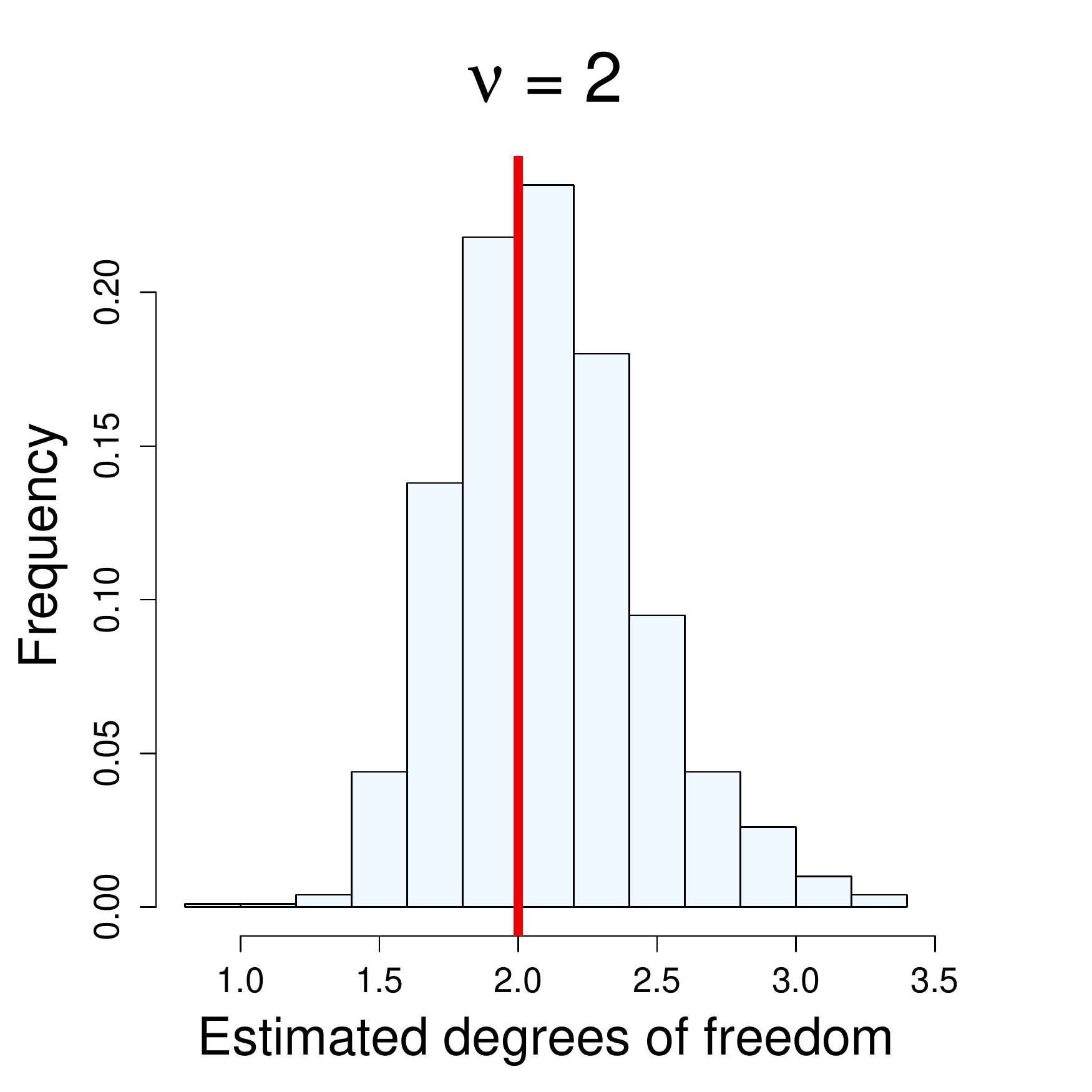}
	\\
	\vspace{1cm}
	\includegraphics[scale=0.4]{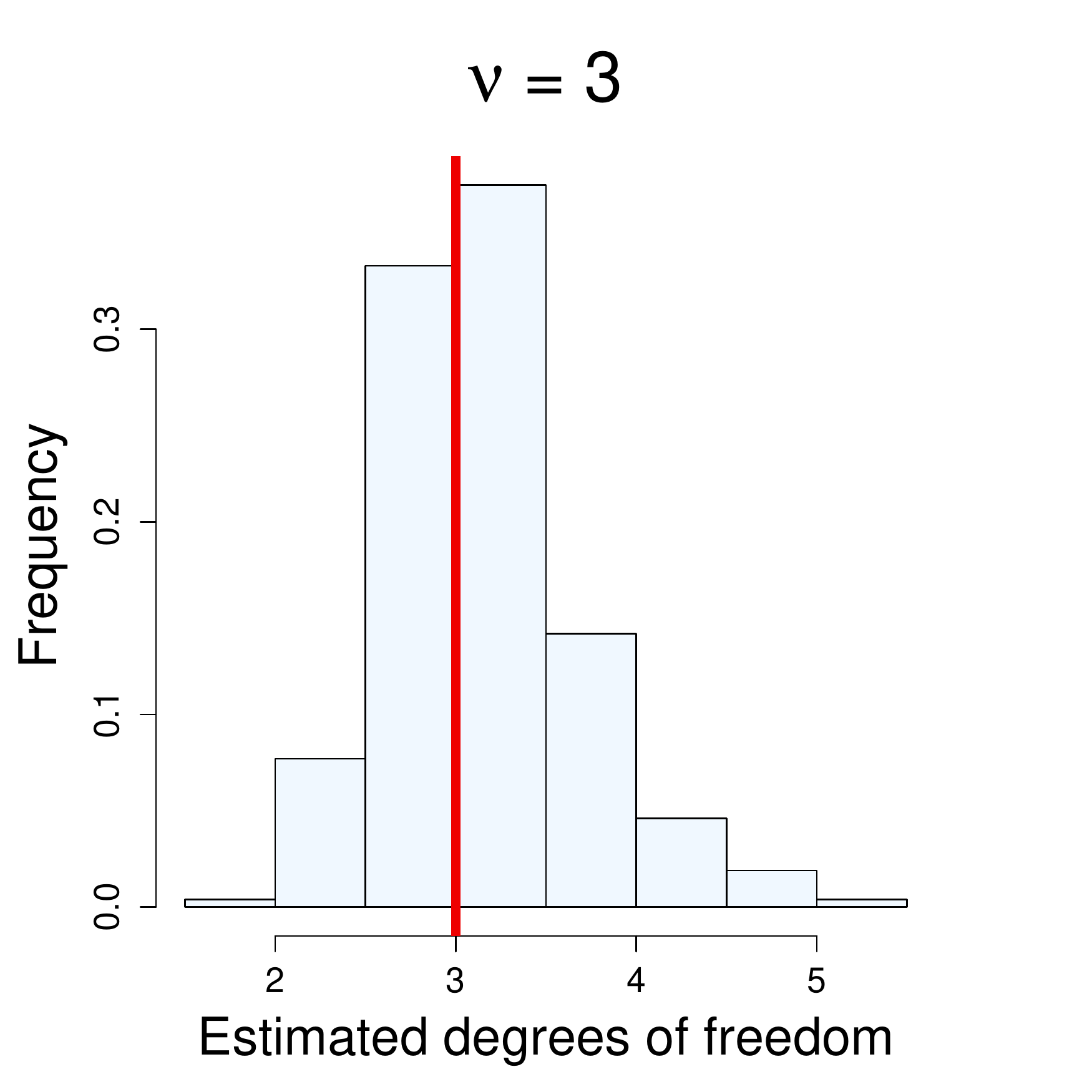} 
	\includegraphics[scale=0.4]{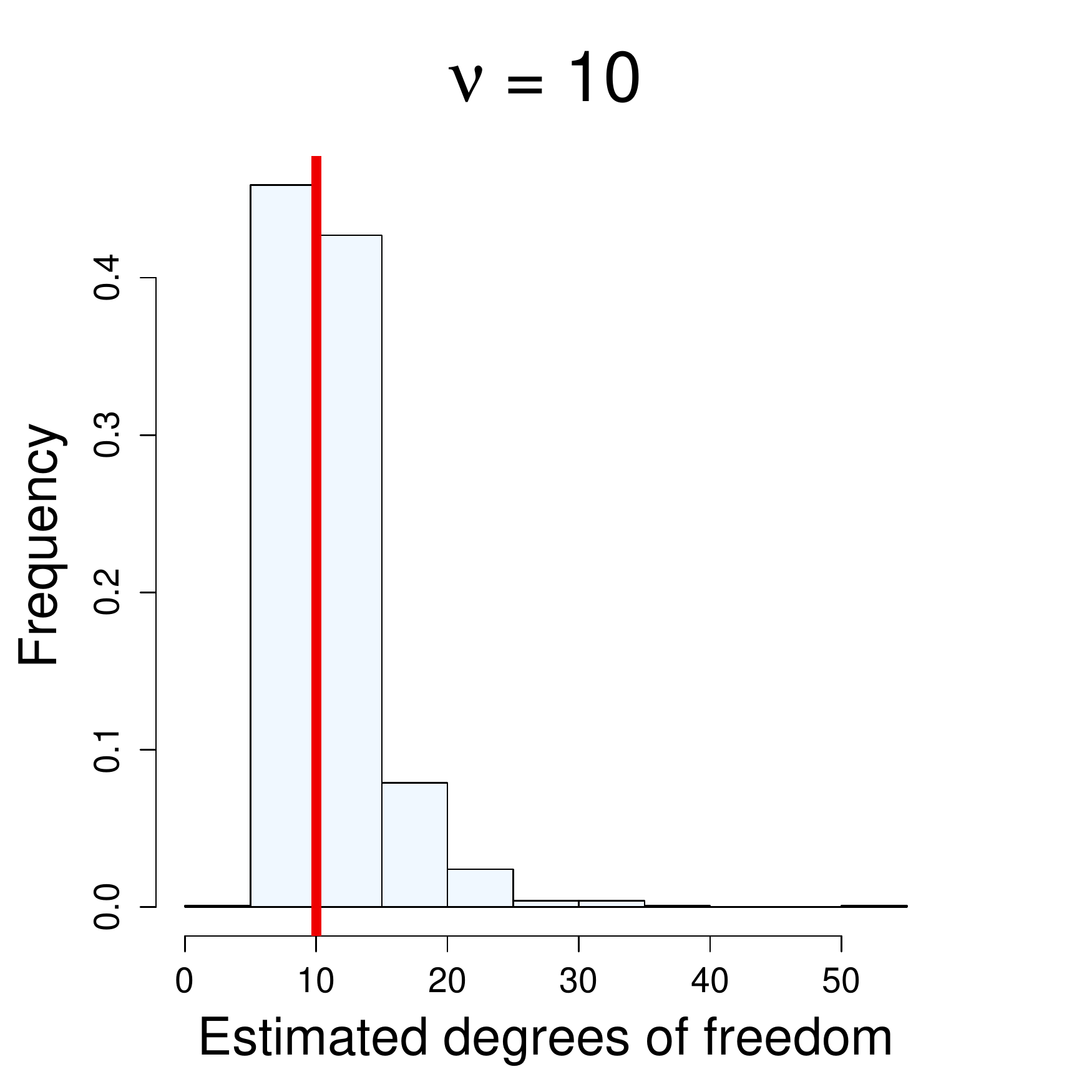}  
	\caption{Frequency of the estimated degrees of freedom,  for  setting $\nu=1,2,3,10$ in the simulation study. The true value of the degrees of freedom is indicated by the vertical red line (dark gray on a gray scale).\label{VOL_df_figures}}
\end{figure}

Finally, Figure \ref{VOL_df_figures} shows the frequencies of the estimated degrees of freedom by the $t$-Lasso, for the different settings $\nu \in \{1, 2, 3, 10\}$. The estimated degrees of freedom are closely centered around the true value (vertical red line). 
The average (averaged over all simulation runs) estimated degrees of freedom are 0.94 (for $\nu=1$), 2.10 (for $\nu=2$), 3.16 (for $\nu=3$) 
 and  11.14 ($\nu=10$).
The variance of the estimated degrees of freedom  increases for larger values of $\nu$.
We conclude that the degrees of freedom are quite accurately estimated. 

\section{Data and volatility spillover analysis \label{VOL_Network_analysis}}

In this section, we first present the data. Then, we define volatility spillovers based on the forecast error variance decomposition. Finally, we present a network analysis tool visualizing these volatility spillovers. 

\subsection{Data\label{VOL_data}}
We study $J=10$ agricultural (corn, wheat, soybean, sugar, cotton, coffee), energy (crude oil, gasoline, natural gas) and biofuel (ethanol) commodities. 
We use daily information about the opening, highest and lowest prices of the nearest to maturity contracts traded in the corresponding future markets. 
Data are available on Thomson Reuters Eikon.
The time span ranges from January $3^{rd}$ 2012 to October $28^{th}$ 2016, thus $T=1070$ daily observations. 
Table \ref{VOL_description} reports the label and the description of each variable.
We obtain univariate volatility measures using the realized daily high-low range proposed by \cite{Parkinson80}, see Appendix
B 
for more details. 
We check for stationarity of the estimated volatilities with univariate unit root tests and the pooled unit root test of \cite{Levin02} and find strong evidence in favor of stationarity ($p\text{-values}<0.01$).


\begin{table}
	\centering
	\caption{Variable description as in Thomson Reuters Eikon for Future Continuation 1 commodities.\label{VOL_description}}
	\medskip
	\begin{tabular}{c|c| l}
		\hline
		\textit{Label} & \textit{Commodity} &\textit{Description} \\
		\hline 
		CRUO & Crude oil & NYMEX Light Sweet Crude Oil (WTI) Composite Energy  \\
		GASO & Gasoline & RBc1 NYMEX RBOB Gasoline Composite Energy \\
		NATG & Natural gas & NGc1 NYMEX Henry Hub Natural Gas Composite Energy\\
		ETHA & Ethanol & CBoT Denatured Fuel Ethanol Electronic Energy \\
		CORN & Corn & Cc1 CBoT Corn Composite Commodity \\
		WHEA & Wheat & Wc1 CBoT Wheat Composite Commodity \\
		SOYB & Soybeans & Sc1 CBoT Soybeans Composite Commodity \\
		SUGA & Sugar & ICE-US Sugar No. 11 Futures Electronic Commodity  \\
		COTT & Cotton & ICE-US Cotton No. 2 Futures Electronic Commodity \\
		COFF & Coffee & KCc1 ICE-US Coffee C Futures Electronic Commodity \\
		\hline
	\end{tabular}
\end{table}

Second, we employ a rolling window over the period 2012-2016 with window size $W=220$ days.
At each time  point $t=W, \ldots, T$, we estimate a VAR($P$) model, with $P$ selected using the Bayesian Information Criterion BIC, for the the log-volatilities from the date in the time window $t-W+1$ until $t$. 
We assume that the error terms follow a multivariate $t$-distribution and estimate the degrees of freedom using the $t$-Lasso.
Then, we study volatility spillovers based on forecast error variance decomposition, as described in the following subsection.

\subsection{Spillover indices \label{VOL_fevd}}
We compute the volatility spillovers by means of the generalized forecast error variance decomposition, taking over the definitions from \cite{Diebold15}.
Consider the Vector Moving Average (VMA) representation of the VAR($P$) in model \eqref{VOL_VARp}
\begin{equation}
\boldsymbol{y}_{t}=\sum_{p=0}^{\infty} \boldsymbol{\theta}_{p} \boldsymbol{e}_{t-p}, \nonumber
\end{equation}
where $\boldsymbol{\theta}_{p}$ is the  moving average coefficient matrix at lag $p$ (cfr. Wold's representation theorem, \citealp{Lutkepohl05}, p25).
Let $\widehat{\boldsymbol{y}}_{t+h}$ be the $h$-step ahead forecast for time $t+h$ made at time $t$ with forecast horizon $h$. 
Then, the $h$-step-ahead forecast error for the $j^{th}$ component of $\boldsymbol{y}_{t}$ is
\begin{equation}
\widehat{y}_{t+h,j} - y_{t+h,j} = \sum_{p=0}^{h-1}
\theta_{p,j1}e_{t+h-p,1} + \ldots + \theta_{p,jJ}e_{t+h-p,J}
,\label{VOL_forecast_error}
\end{equation}
where $e_{t,j}$ is the $j^{th}$ component of $\boldsymbol{e}_{t}$, and $\theta_{p,jk}$ is the $jk^{th}$ entry of $\boldsymbol{\theta}_{p}$.

If  an impulse to  $e_{t,k}$ of size of one standard deviation is given, then the expected value of the error term equals
\begin{equation}
	E(\boldsymbol{e}_t|e_{t,k}=\sqrt{\sigma_{kk}})=\dfrac{\boldsymbol{\Sigma}\boldsymbol{\delta}_k}{\sqrt{\sigma_{kk}}}, \label{VOL_conditionall_exp_shock}
\end{equation}
where $\sigma_{kk}$ is the $kk^{th}$ entry of $\boldsymbol{\Sigma}$ and $\boldsymbol{\delta}_k$ is the selection vector of length $J$ with unity entry as its $k^{th}$ element and zeros elsewhere.
Equation \eqref{VOL_conditionall_exp_shock} defines a generalized impulse, as in \cite{Pesaran98}, with response vector $\boldsymbol{\theta}_p\boldsymbol{\Sigma}\boldsymbol{\delta}_k/\sqrt{\sigma_{kk}}$, for $p=1,\ldots,P$.
It is important to note that \eqref{VOL_conditionall_exp_shock} does not only hold for a normal distribution, but also for a $t$-distribution (e.g. \citealp{Ding16}).
If $\nu \leq 2$ and the covariance matrix of the $t$-distribution is not existing, the $\boldsymbol{\Sigma}$ in \eqref{VOL_conditionall_exp_shock} should be replaced by the $\boldsymbol{\Psi}$ scale matrix.

The $jk^{th}$ entry of the $h$-step ahead variance decomposition matrix is then defined as
\begin{equation}
	w_{h,jk}= \dfrac{\sigma_{kk}^{-1}\sum_{p=0}^{h-1}
		(\boldsymbol{\delta}_{j}'\boldsymbol{\theta}_{p}\boldsymbol{\Sigma}\boldsymbol{\delta}_{k})^{2}}
	{\sum_{p=0}^{h-1}\boldsymbol{\delta}_{j}'\boldsymbol{\theta}_{p}\boldsymbol{\Sigma}\boldsymbol{\theta}_{p}'\boldsymbol{\delta}_{j}}.\label{VOL_GFEVD}	
\end{equation}
This is the proportion of the variance of the $h$-step-ahead forecast error \eqref{VOL_forecast_error} that is accounted for by the innovations in variable $k$ of the VAR.  
Note that the denominator of \eqref{VOL_GFEVD} equals the variance of \eqref{VOL_forecast_error}, and the numerator of \eqref{VOL_GFEVD} is the squared $j^{th}$ component  of the response vector.
As the error term components are not orthogonal, in general $\sum_{k=1}^{J} w_{h,jk} \neq 1$. We consider the normalized variance decomposition
\[
\tilde{w}_{h,jk}=\dfrac{w_{h,jk}}{\sum_{k=1}^{J} w_{h,jk}},
\]
which is in the interval $[0,1]$  by construction. The above normalized variance decomposition corresponds to the generalized variance decomposition proposed by \cite{Lanne16}.

The \textit{volatility spillover} from commodity $k$ to commodity $j$ is defined as
\begin{equation}
s_{h,k \rightarrow j} = 100* \tilde{w}_{h,jk}. \label{VOL_spillovers}
\end{equation}
The \textit{volatility spillover index} is given by 
\begin{equation}
s_{h}=\sum_{{\substack{ j\neq k \\ j,k=1 }}}^{J} s_{h,k \rightarrow j}, \label{VOL_spilloverindex}
\end{equation}
as a single measure of the overall volatility spillovers and  an overall proxy of the magnitude of the volatility spillovers.
In the remainder, we take $h=5$, which coincides to a working week \citep{Bubak11}.

\subsection{Network analysis\label{VOL_network}}
We visualize the volatility spillovers using a network analysis tool (e.g. \citealp{Diebold15, Hautsch15}).
The nodes in the network are the different commodities. An edge from commodity $k$ to commodity $j$ is drawn if $s_{h,k \rightarrow j}$ from equation \eqref{VOL_spillovers} is non-zero; the edge width represents the magnitude of the volatility spillover. Hence, our network is directed and weighted, however, not necessarily sparse. Indeed, the sparsity of $\widehat{\boldsymbol{B}}$ is not necessarily preserved in the estimated VMA coefficient matrices $\widehat{\boldsymbol{\theta}}$.
Consequently, the network will picture many non-zero volatility spillovers and hence contain many edges, although many of them may be quite small. 
Therefore, for better visualization, we give the network representing only the $15\%$ largest volatility spillovers. 

%

\section{Results \label{VOL_DataResults}} 	
 
In this section, we first present the time evolution of the estimated degrees-of-freedom  and of the volatility spillovers (cfr. Section \ref{VOL_data}). Second, we picture the networks showing the volatility spillovers (cfr. Section \ref{VOL_network}). 
Finally, we show the good performance of the $t$-Lasso in terms of forecast accuracy.   

\subsection{Rolling window analysis}

\begin{figure}
	\centering
	\includegraphics[scale=0.38]{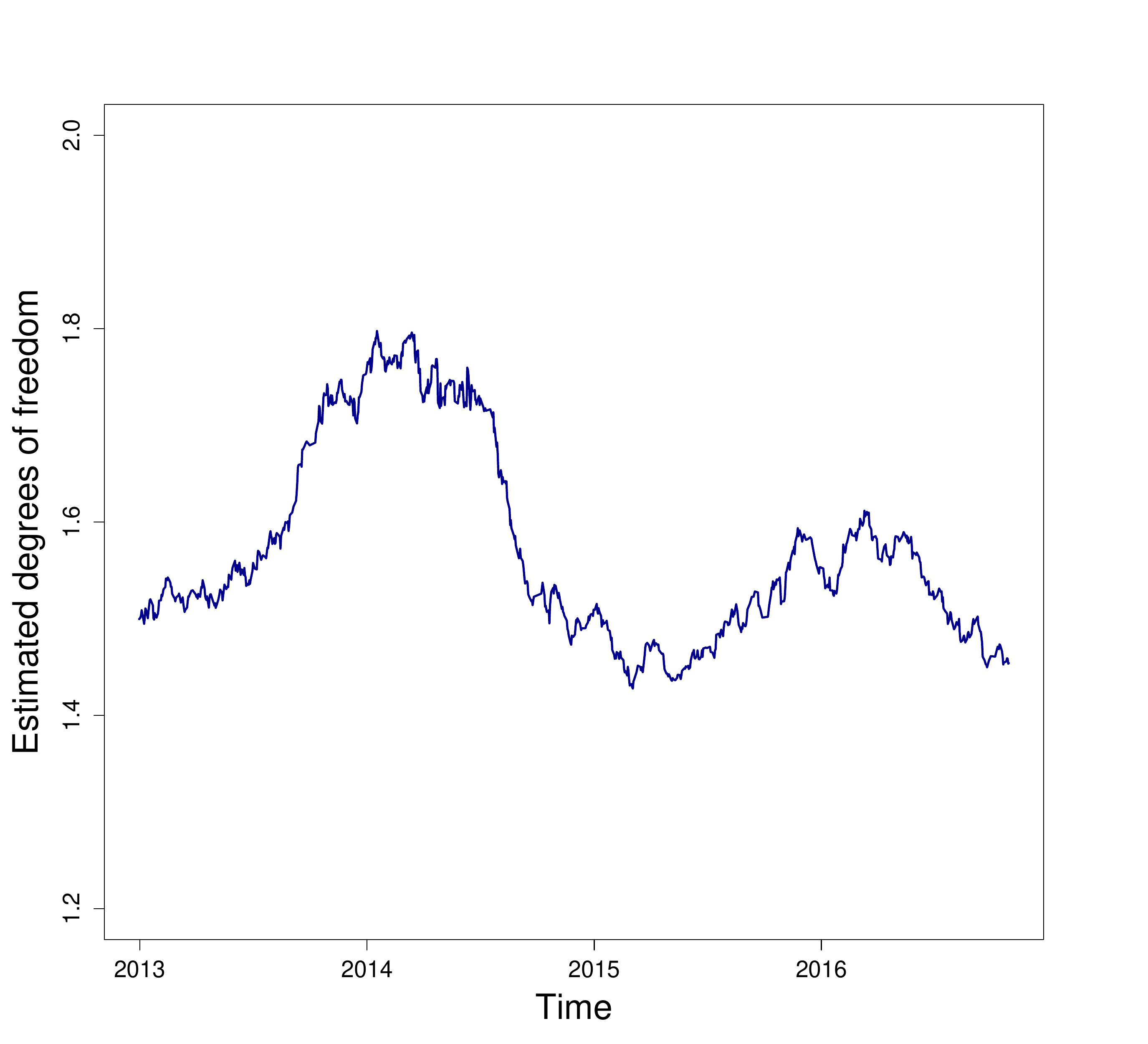}
	\caption{Estimated degrees of freedom of the multivariate $t$-distribution of the VAR residuals by the $t$-Lasso for a 220-day rolling window. 
	The $x$-axis represents the ending date of each window. \label{VOL_estimated_df} }
\end{figure}

\paragraph{Estimated degrees of freedom.}
Figure \ref{VOL_estimated_df} reports the estimated degrees of freedom
of the multivariate $t$-distribution of the VAR residuals at each time point $t$, with $t$ the end point of each time window. 
The average value of $\widehat{\nu}$ is 1.57, with maximum 1.80 and minimum 1.43. 
Overall, we observe that the estimated degrees of freedom are very low.
This confirms the existence of heavy tails and justifies the use of the $t$-Lasso rather than the Gaussian  Lasso.

If we look at the evolution of $\widehat{\nu}$ over time, we detect lower values of the estimated degrees of freedom in the time windows ending between the second half of 2014 and the first months of 2015.
Volatile commodity markets characterized the second half of 2014, mainly driven by the drop of almost 50\% in crude oil price \citep{Knittel16}.
Smaller values of $\widehat{\nu}$ indicate more pronounced extreme realizations in the VAR innovations and reflect a less predictable behavior of commodity markets. 

\begin{figure}
	\centering
	\includegraphics[scale=0.38]{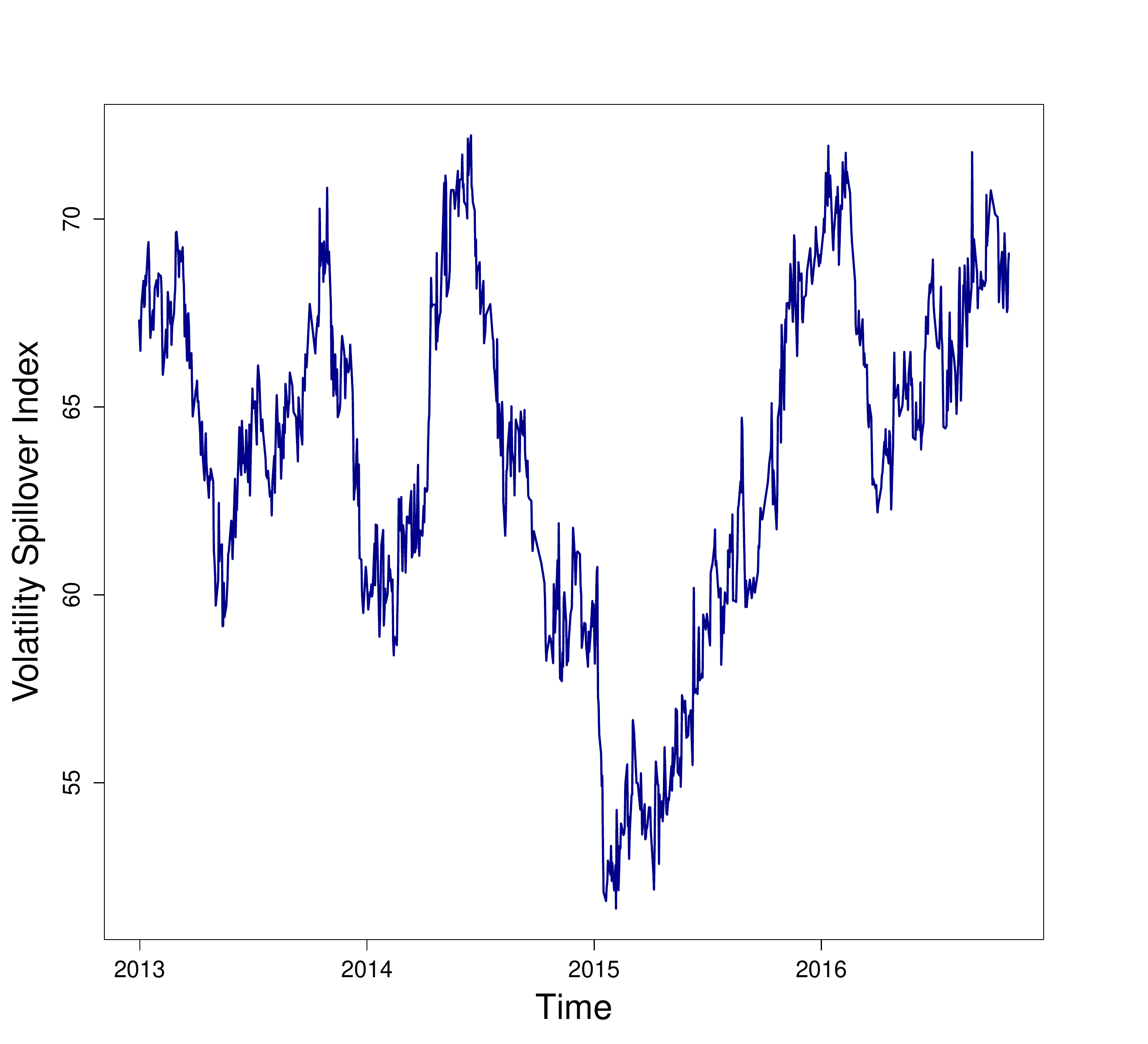}
	\caption{Volatility spillover index for the $t$-Lasso for a 220-day rolling window. The $x$-axis represents the ending date of each window. \label{VOL_rolling_global_spill}}
\end{figure}

\paragraph{Volatility spillover index.}
Figure \ref{VOL_rolling_global_spill} reports the evolution of the volatility spillover index (cfr. equation \eqref{VOL_spilloverindex}) for the $t$-Lasso as a function of time $t$, being $t$ the end point of each time window.
We observe that the volatility spillover index experiences large swings over time. In particular, in the second half of 2014 we detect a large drop in the overall level of volatility spillovers. This drop is not permanent and during 2015 the volatility spillover index returns to the level prior to 2014.

The downturn in volatility spillover index can also be linked to the fall of crude oil price that occurred in the second half of 2014. 
Crude oil price experienced large downturns driven by (i) a larger supply from OPEC and non-OPEC countries and (ii) a weak global demand due to the slowdown of the world economy, notably the Chinese one. 
The drop in energy prices made biofuel less profitable and implied lesser and weaker volatility spillovers among energy, biofuel and agriculture commodities. 
 
\subsection{Network analysis}
\begin{figure}
	\centering
		\includegraphics[scale=0.55]{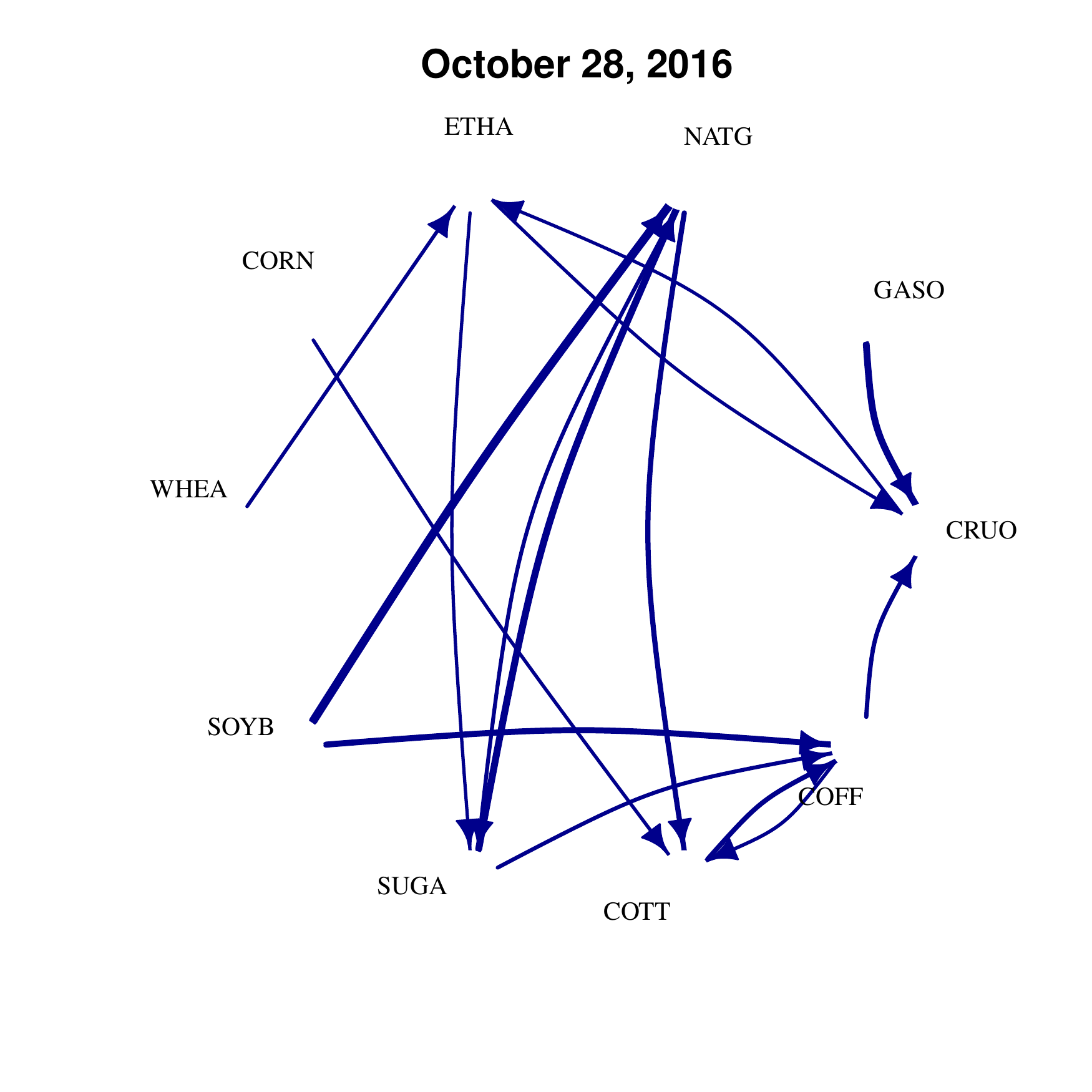} \\
		\hspace{-2cm}
		\includegraphics[scale=0.55]{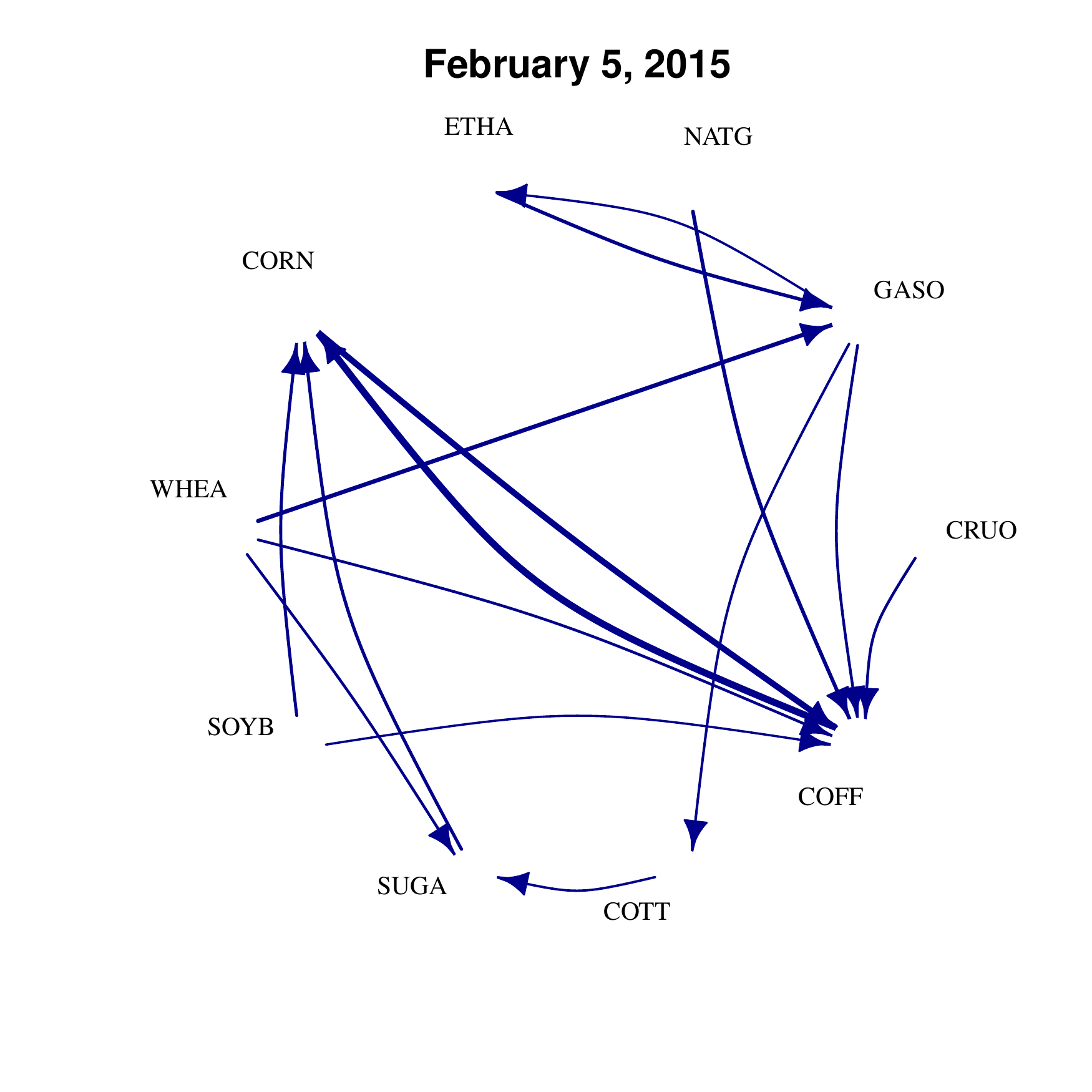}  \hspace{-1.5cm}
		\includegraphics[scale=0.55]{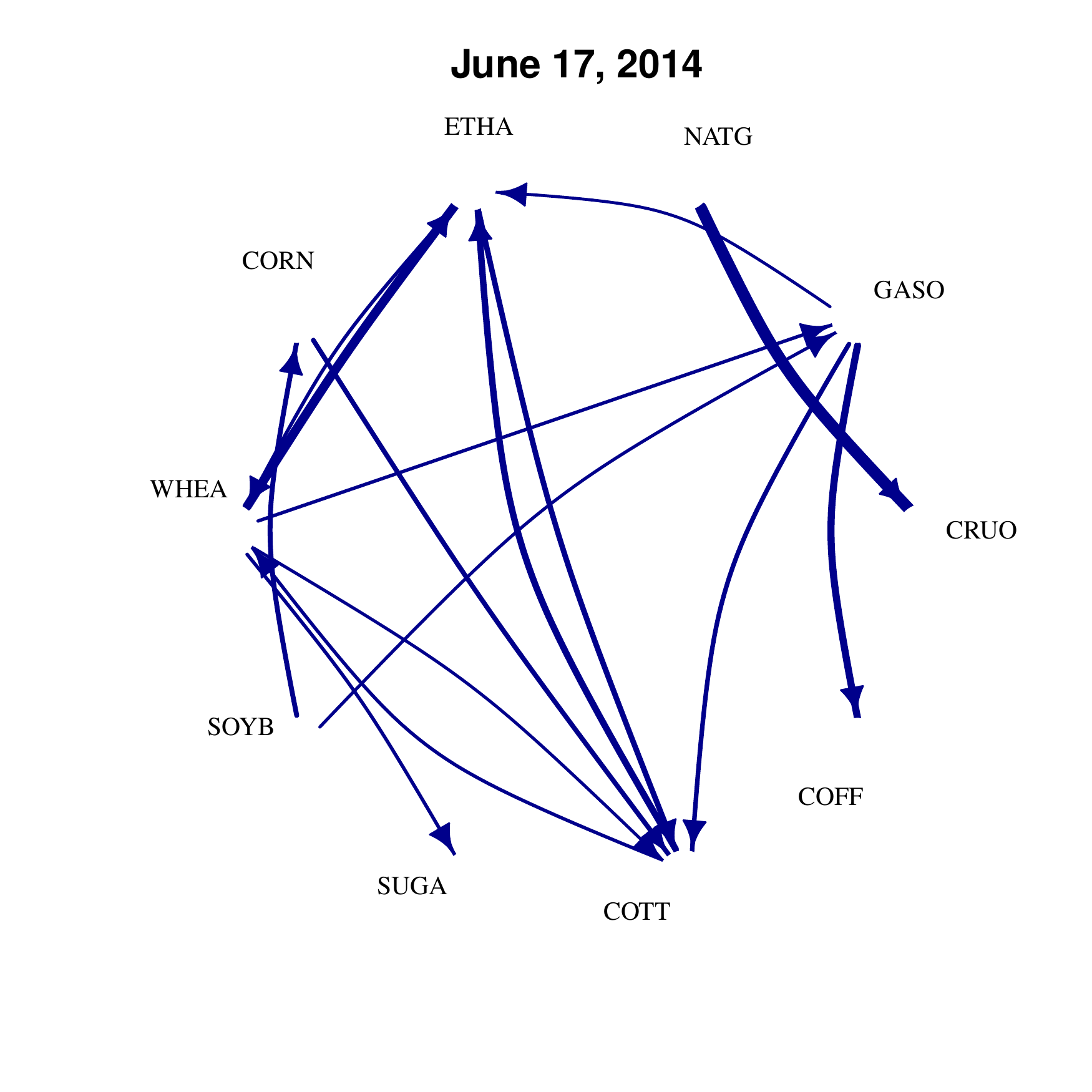} 
		\hspace{-1cm}
		
		\caption{Commodity networks of volatility spillovers for different ending times of the $220$-day rolling windows: October $28^{th}$ 2016 (top), February $5^{th}$ 2015 (bottom left) and June $17^{th}$ 2014 (bottom right). 
		}\label{Volatility_networks}
\end{figure}

We visualize the volatility spillover in a network of commodities.
Figure \ref{Volatility_networks} top presents the network for the time window ending on October $28^{th}$ 2016, the last time point in our data set.
For instance, a directed edge is drawn from gasoline (GASO) to crude oil (CRUO)  since the volatility spillover from gasoline towards crude oil is different from zero and belongs to the largest 15\% volatility spillovers.
Recall that the edge width represents the magnitude of the volatility spillover.
Figure \ref{Volatility_networks} also reports the networks for the time windows ending on February $5^{th}$ 2015 (bottom left) and on June $17^{th}$ 2014 (bottom right). For simplicity, from now on we refer to the three networks only by the year of the ending date of their windows. 
The 2015 and 2014 networks correspond to the time windows with the lowest and highest, respectively, values of the volatility spillover index.

\paragraph{Link energy-biofuel.}
In all networks we find volatility spillovers among energy and biofuel commodities.
In the 2015 and 2014 networks, gasoline is the only energy commodity connected by an edge with ethanol.
As ethanol is often blended in gasoline for consumption \citep{Serra13}, it is not surprising to observe volatility spillovers between the two commodities in times of high energy prices (i.e. network 2014) and in times of large energy price changes (i.e. network 2015).  

\paragraph{Link energy-agriculture.}
Bidirectional volatility spillovers between energy and agriculture are detected in all networks confirming the findings of \cite{Rezitis15}. 
In the 2016 network, natural gas shows numerous and large spillovers from and towards agricultural commodities, whereas in the 2015 and 2014 networks, gasoline is the most connected energy commodity with agriculture.
Both natural gas and gasoline are key commodities in the agricultural sector: the former is the major production cost for fertilizers, whereas the latter is the most important fuel (together with diesel).
If energy prices are high (e.g. network 2014), fuel has a large impact on agricultural commodities, which could explain the numerous volatility spillovers from/to gasoline.
Conversely, as energy prices are low and fuel is cheap (e.g. network 2016), variations in fertilizer prices  largely affect agricultural commodities volatilities, which could explain the many volatility spillovers from/to natural gas.

In all networks, various agricultural commodities are involved in volatility spillovers to and from energy.
In general, volatility spillovers from and to energy involve both agricultural crops that are primary inputs for biofuel production - like wheat, soybean and sugar -, and crops that do not have a direct link with biofuels - like coffee -, as in \cite{Rezitis15}.

\paragraph{Link biofuel-agriculture.}
Volatility spillovers between biofuel and agriculture are found only in the 2016 and 2014 networks.
These volatility spillovers are bidirectional and involve wheat, sugar and cotton: this is consistent with our expectation since these crops can be used for ethanol production.
In the 2015 network, we observe no volatility spillovers between biofuel and agriculture. In that period energy prices substantially dropped \citep{Knittel16}: this made biofuel less attractive compare to other standard fuels and could have resulted in weaker volatility spillovers from and to agriculture.

\subsection{Forecast accuracy}
We now  forecast the commodity volatilities. 
For a given forecast horizon, we forecast the vector of log volatilities  $\widehat{\boldsymbol{y}}_{t+h}^{(t)}=[\widehat{y}_{t+h,1}^{(t)}, \ldots, \widehat{y}_{t+h,J}^{(t)}]$ based on the time window $[t-W+1,t]$, where $t$ is the end point of the time window of size $W$ and $\widehat{y}_{t+h,j}^{(t)}$ is the $h$-step ahead forecast of series $j$ at time $t+h$ made at this end point.  
The forecast is obtained using the VAR model in \eqref{VOL_VARp}, with autoregressive coefficients $\widehat{\boldsymbol{B}}$ obtained with the $t$-Lasso with $\nu$ estimated, the Gaussian Lasso and the LS.
We compare these three estimators in terms of forecast accuracy by computing the Mean Absolute Forecast Error
\begin{equation*}
\text{MAFE}=
\dfrac{1}{N-h-W+1}
\sum_{t=W}^{N-h} \text{MAFE}_t
\ \ \
\text{with}
\ \ \
\text{MAFE}_t=
\dfrac{1}{J}
\sum_{j=1}^{J}
|\widehat{y}_{t+h,j}^{(t)} - y_{t+h,j}|. \label{VOL_MAFE}
\end{equation*}
Hence, the MAFE is computed for each time window and then averaged across all of them. 
The smaller the value of the mean absolute forecast error, the more accurate the volatility forecasts.

\paragraph{Results.} 
Table \ref{VOL_MAFElog_tab} reports the results of the MAFE for the three estimators for different forecast horizons, namely $h=1$, $h=5$ and $h=20$ days.  
The $t$-Lasso attains the best value of the MAFE for all forecast horizons. 
We find that based on the Diebold-Mariano test \citep{Diebold95}, the difference in forecast accuracy with the other estimators is always significant ($p{\text{-values}}<0.01$).
For forecast horizon $h=1$, the $t$-Lasso gives a relative improvement in MAFE of 1\% and of 23\% over the Gaussian Lasso and the LS, respectively. 
The higher the forecast horizon, the better the performance of the $t$-Lasso compared to the Gaussian Lasso: for instance, for $h=20$ our estimator gives an improvement of  14\%. 

\begin{table}
	\centering
	\caption{Mean Absolute Forecast Error for the $t$-Lasso, the Gaussian Lasso and the LS for    forecast horizons $h=1,5$ and 20 from a 220-day rolling window.\label{VOL_MAFElog_tab}}
	\medskip
	\begin{tabular}{l|ccc}
		\hline
		Horizon & $t$-Lasso & Gaussian Lasso & LS \\
		\hline
		$h=1$ & 1.871 & 1.896 & 2.418 \\ 
		$h=5$ & 1.959 & 2.056 & 2.589 \\ 
		$h=20$ & 2.280 & 2.636 & 2.820 \\
		\hline
	\end{tabular}
\end{table}

\begin{figure}
	\centering
	\includegraphics[scale=0.44]{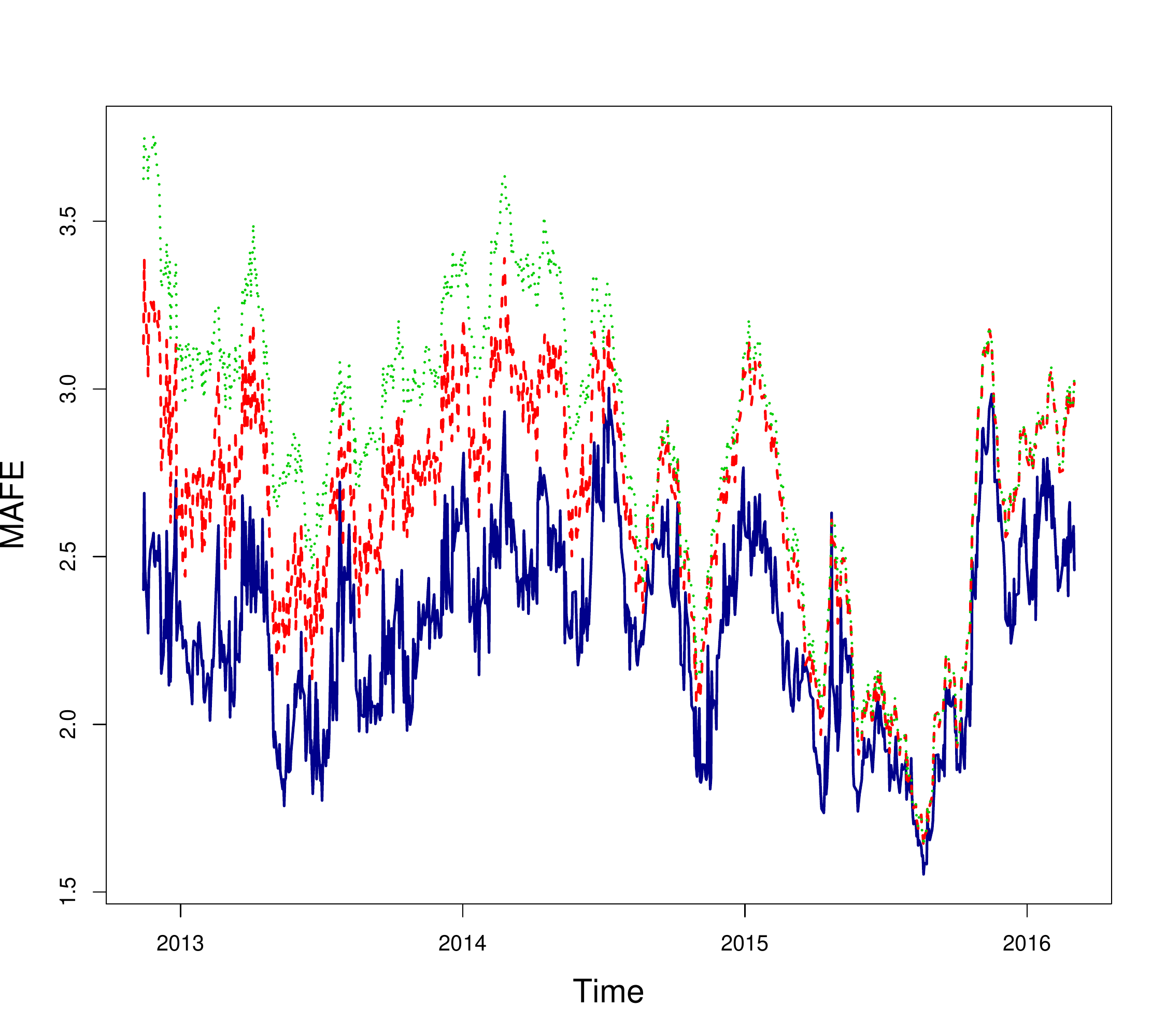}	
	\caption{Mean Absolute Forecast Error for a 220-day rolling window and forecast horizon $h=20$ for the $t$-Lasso (solid), the Gaussian Lasso (dashed) and the Leas Squares (dotted).
		The $x$-axis represents the ending date of each window. \label{VOL_MAFE_roll}}
\end{figure}

Figure \ref{VOL_MAFE_roll} reports the evolution of the MAFE$_t$ (for $h=20$) for the $t$-Lasso (blue solid line), the Gaussian Lasso (red dashed line) and the Least Squares (green dotted line) as a function of $t$, the end point of each time window. 
The $t$-Lasso attains a lower mean absolute forecast error than the Gaussian Lasso in all but five time windows and the difference in forecast accuracy is confirmed to be significant in all these time windows by the Diebold-Mariano test. 
Overall, we find that the $t$-Lasso attains a better forecast performance than the Gaussian Lasso and the LS, regardless of the forecast horizon or the presence of larger volatility spillovers.
 
\section{Discussion\label{VOL_Discussion}}
This paper studies volatility spillovers in a Vector AutoRegressive (VAR) model accounting for the fat-tailedness of its innovations. 
We extend the work of \cite{Diebold15} and propose a penalized estimator of the VAR model with errors following a multivariate $t$-distribution, of which we include the estimation of the degrees of freedom.
We emphasize that the normal distribution is a special case, for $\nu=\infty$.
Our estimator is computable even for a large number of relatively short time series. 
Our simulation study shows that the proposed $t$-Lasso attains a better performance in terms of estimation accuracy than the Gaussian Lasso or Least Squares. 
The penalized estimator ensures good estimation accuracy even for VAR models that contain a large number of time series relative to the time series length.
The $t$-distribution for the errors better captures the spikes typical of commodity volatility.

We study the dynamics of volatility spillovers between $J=10$ energy, biofuel and agricultural commodities using a rolling window approach.
Our findings highlight that the distributions of the log volatilities and of the residuals of the VAR model are fat-tailed. 
Furthermore, we visualize volatility spillovers using networks built on forecast error variance decomposition.
We find evidence of bidirectional volatility spillovers between energy and agricultural commodities, regardless of the fact of being biofuel crops or not \citep{Rezitis15}. 

Our analysis is built on the volatility spillover definition by \cite{Diebold14} and relies on generalized forecast error variance decomposition. Nevertheless, other definitions could have been used. 
For instance, we redid the analysis using Cholesky's decomposition (\citealp{Diebold15}, p14) and the spectral decomposition \citep{Hafner06}. 
In both cases, we obtain results comparable to the ones reported, with only minor changes in the magnitudes of the volatility spillovers.
The same holds when using other volatility spillover definitions. For instance, we redid the analysis defining a volatility spillover as the sum of squared generalized impulse responses, and draw comparable conclusions. 
Detailed results are available from the authors upon request.


Further research might follow several trajectories. 
On the one hand, one might investigate the existence of asymmetries in the VAR errors, for instance by considering a multivariate skewed $t$-distribution (\citealp{Kotz04}, p98) or by studying realized semi-variances as in \cite{Barunik15}.
On the other hand, it could be interesting to use other data and volatility measures \citep{McAleer08}. 
For commodity futures, data on the opening, lowest and highest daily prices are easier to obtain than intra-day high-frequency returns, motivating our choice of studying daily realized ranges. 
However, the innovation distribution  could be less or more heavy-tailed depending on the asset type and/or the volatility measure, and this might have different implications in terms of risk management (for instance, see \citealp{Brownlees10} for a Vale-at-Risk comparison between different realized measures of volatility for NYSE stocks).

Finally, the proposed $t$-VAR approach could be used in other research  that  study lagged effects among a large number of time series accounting for fat-tailed errors.
Among other ones, the analysis of systemic risk,  that is the risk that an individual firm default might threaten the stability of an entire market \citep{Hautsch15}, seems a natural application.
After the 2007-09, great importance has been given to the modeling of the spillovers among firms as a measure of systemic risk:
typically, one  would study a large number of firms \citep{Hautsch15}, and model the non-normal innovation distribution caused by extreme downturns in the equity market \citep{Engle15}. 
The proposed $t$-VAR approach would clearly address both issues, 
and its results would be highly interpretable in the form of risk networks.


\bigskip

\noindent
{\bf Acknowledgments.}
This work was supported by the FWO (Research Foundation Flanders, contract number 12M8217N) and by the GOA/12/014 project of the Research Fund KU Leuven.
We thank Piet Sercu and Kris Boudt for the useful comments that greatly improved the manuscript.


\section*{Appendix A: Algorithm Gaussian Lasso\label{VOL_appendix_gaussian}}
The algorithm used to compute the Gaussian Lasso is given in Algorithm 
A
. 
First, we solve for the autoregressive parameter $\boldsymbol{B}$ conditional on the inverse error covariance matrix $\boldsymbol{\Omega}$ using a coordinate descent algorithm, as in \cite{Friedman07}.
Second we solve for  $\boldsymbol{\Omega}$ conditional on $\boldsymbol{B}$, using the Graphical Lasso algorithm of \cite{Friedman08}. 
These two steps are implemented in \texttt{R} using the \texttt{grplasso} and \texttt{glasso} packages,  respectively. 
We iteratively repeat the two steps until convergence of the objective function is reached.

\begin{algorithm}
	\caption{\hspace{-0.15cm}\textbf{A} \ Gaussian Lasso: penalized estimation with Gaussian errors \label{VOL_Algorith_Gaus}}
	\begin{description}
		\item [{Input}] $\textbf{Y}$, $\textbf{X}$,  and desired accuracy $\varepsilon$. 
		\item [{Initialization}] $\widehat{\boldsymbol{\Omega}}^{(0)}=\textbf{I}_{J}$.
		\item[{Iteration}] Iterate the following steps for $m=0,1,2,\ldots$:
		\begin{description}
			\item [{Solving for $\boldsymbol{B}|\boldsymbol{\Omega}$}.] Compute $\widehat{\boldsymbol{B}}^{(m+1)}$:
			\[
			\widehat{\boldsymbol{B}}^{(m+1)}= \underset{\boldsymbol{B}}{\operatorname{argmin}} \ \dfrac{1}{2N} 
			\operatorname{tr}\left[(\boldsymbol{Y} - \boldsymbol{X}\boldsymbol{B})\widehat{\boldsymbol{\Omega}}^{(m)}(\boldsymbol{Y} - \boldsymbol{X}\boldsymbol{B})'\right]
			+ \lambda \sum_{i,j=1}^{J} \sum_{p=1}^{P} |B_{p,ij}|.
			\] 
			
			\item [{Solving for $\boldsymbol{\Omega}|\boldsymbol{B}$}.] Compute $\widehat{\boldsymbol{\Omega}}^{(m+1)}$:
			\[
			\widehat{\boldsymbol{\Omega}}^{(m+1)}= \underset{\boldsymbol{\Omega}}{\operatorname{argmin}} \ \dfrac{1}{2N} 
			\operatorname{tr}\left[(\boldsymbol{Y} - \boldsymbol{X}\widehat{\boldsymbol{B}}^{(m+1)})\boldsymbol{\Omega}(\boldsymbol{Y} - \boldsymbol{X}\widehat{\boldsymbol{B}}^{(m+1)})'\right]
			-\dfrac{1}{2}\log|\boldsymbol{\Omega}|
			+ \gamma \sum_{i \neq j}^{J}|\omega_{ij}|.
			\] 
		\end{description}
		\item [{Convergence}] Iterate until the relative change in the value of the objective function in \eqref{VOL_JGrLasso} in two successive iterations is smaller than $\varepsilon$.
		\item[Output] $\widehat{\boldsymbol{B}}=\widehat{\boldsymbol{B}}^{(m+1)}$ and  $\widehat{\boldsymbol{\Omega}}=\widehat{\boldsymbol{\Omega}}^{(m+1)}$.
	\end{description}
\end{algorithm}

\paragraph{Selection of regularization parameters.}	
When solving for $\boldsymbol{B}|\boldsymbol{\Omega}$, we use a grid of regularization parameters $\lambda$ and search for the optimal one minimizing the Bayesian Information Criterion (BIC)
\[
\text{BIC}_{\lambda} = - 2 \text{log} L_{\lambda} + \text{df}_{\lambda} \text{log}(N),
\] 
where $\text{log} L_{\lambda}$ is the estimated likelihood, i.e. the first term in  \eqref{VOL_JGrLasso}, using regularization parameter $\lambda$, and $\text{df}_{\lambda}$ is the number of non-zero components of $\widehat{\boldsymbol{B}}_{\lambda}$.
Likewise, when solving for  $\boldsymbol{\Omega}|\boldsymbol{B}$ we use a grid of regularization parameters $\gamma$ and search for the optimal one minimizing the BIC
\[
\text{BIC}_{\gamma} = - 2 \text{log} L_{\gamma} + \text{df}_{\gamma} \text{log}(N),			
\]
where $df_{\gamma}$ is the number of non-zero lower diagonal elements of $\widehat{\boldsymbol{\Omega}}$.

\section*{Appendix B: Volatility measure \label{VOL_appendix_ranges}}

Following \cite{Parkinson80}, we obtain a measure of volatility of a future contract using the high-low daily range estimator. Consider the daily information about opening price $O_{t,j}$, the highest price $H_{t,j}$ and the lowest price $L_{t,j}$ attained on date $1 \leq t \leq T$ for commodity $1 \leq j \leq J$. 
The high-low range estimator for the daily variance is
\[
v_{t,j}=\dfrac{1}{4\log(2)}(h_{t,j} - l_{t,j})^2,
\]
where $h_{t,j}=\log(H_{t,j}) - \log(O_{t,j})$ and $l_{t,j}=\log(L_{t,j}) - \log(O_{t,j})$ are the maximum and minimum daily return, respectively.
For a review on realized range measures of volatility see \cite{Shu06} or \cite{Martens07}.
The time series entering the VAR of equation \eqref{VOL_VARp} are the log trasformations of the realized ranges $\widehat{v}_{t,j}$ of commodity $j$ at time $t$, that is ${\bf y}_t=[\log(\widehat{v}_{t,j}),\ldots,\log(\widehat{v}_{t,J})]'$. 
The original volatility series can be obtained by applying the exponent transformation and correction for the re-transformation bias as in \cite{Bauer11}.

\bibliographystyle{asa}
\bibliography{Volatility_ref}

\begin{thebibliography}{44}
\newcommand{\enquote}[1]{``#1''}
\expandafter\ifx\csname natexlab\endcsname\relax\def\natexlab#1{#1}\fi

\bibitem[{Andersen et~al.(2001)Andersen, Bollerslev, Diebold, and
  Heiko}]{Andersen01ST}
Andersen, T.~G.; Bollerslev, T.; Diebold, F.~X. and Heiko, E. (2001),
  \enquote{{The distribution of realized stock return volatility},}
  \textit{Journal of Financial Economics}, 61(1), 43--76.

\bibitem[{Barigozzi and Hallin(2017)}]{Barigozzi17}
Barigozzi, M. and Hallin, M. (2017), \enquote{{A network analysis of the
  volatility of high dimensional financial series},} \textit{Journal of the
  Royal Statistical Society: Series C (Applied Statistics)}, 66(3), 581--605.

\bibitem[{Barndorff-Nielsen and Shephard(2002)}]{BarndorffNielsen02}
Barndorff-Nielsen, O.~E. and Shephard, N. (2002), \enquote{{Econometric
  analysis of realized volatility and its use in estimating stochastic
  volatility models},} \textit{Journal of the Royal Statistical Society: Series
  B (Statistical Methodology)}, 64(2), 253--280.

\bibitem[{Barun\'{i}k et~al.(2015)Barun\'{i}k, Ko\v{c}enda, and
  V\'{a}cha}]{Barunik15}
Barun\'{i}k, J.; Ko\v{c}enda, E. and V\'{a}cha, L. (2015), \enquote{{Volatility
  spillovers across petroleum markets},} \textit{The Energy Journal}, 36(3),
  309--330.

\bibitem[{Basu and Michailidis(2015)}]{Basu15}
Basu, S. and Michailidis, G. (2015), \enquote{Regularized estimation in sparse
  high-dimensional time series models,} \textit{The Annals of Statistics},
  43(4), 1535--1567.

\bibitem[{Bauer and Vorkink(2011)}]{Bauer11}
Bauer, G.~H. and Vorkink, K. (2011), \enquote{{Forecasting multivariate
  realized stock market volatility},} \textit{Journal of Econometrics}, 160(1),
  93--101.

\bibitem[{Brownlees and Gallo(2010)}]{Brownlees10}
Brownlees, C.~T. and Gallo, G.~M. (2010), \enquote{{Comparison of volatility
  measures: A risk management perspective},} \textit{Journal of Financial
  Econometrics}, 8(1), 29--56.

\bibitem[{Bub{\'a}k et~al.(2011)Bub{\'a}k, Ko{\v{c}}enda, and
  {\v{Z}}ike{\v{s}}}]{Bubak11}
Bub{\'a}k, V.; Ko{\v{c}}enda, E. and {\v{Z}}ike{\v{s}}, F. (2011),
  \enquote{Volatility transmission in emerging European foreign exchange
  markets,} \textit{Journal of Banking and Finance}, 35(11), 2829--2841.

\bibitem[{Callot et~al.(2017)Callot, Kock, and Medeiros}]{Callot17}
Callot, L. A.~F.; Kock, A.~B. and Medeiros, M.~C. (2017), \enquote{{Modeling
  and forecasting large realized covariance matrices and portfolio choice},}
  \textit{Journal of Applied Econometrics}, 32(1), 140--158.

\bibitem[{Caporin and Velo(2015)}]{Caporin15}
Caporin, M. and Velo, G.~G. (2015), \enquote{{Realized range volatility
  forecasting: Dynamic features and predictive variables},}
  \textit{International Review of Economics and Finance}, 40, 98--112.

\bibitem[{Christensen et~al.(2009)Christensen, Podolskij, and
  Vetter}]{Christensen09}
Christensen, K.; Podolskij, M. and Vetter, M. (2009), \enquote{{Bias-correcting
  the realized range-based variance in the presence of market microstructure
  noise},} \textit{Finance and Stochastics}, 13(2), 239--268.

\bibitem[{Corsi et~al.(2008)Corsi, Mittnik, Pigorsch, and Pigorsch}]{Corsi08}
Corsi, F.; Mittnik, S.; Pigorsch, C. and Pigorsch, U. (2008), \enquote{{The
  volatility of realized volatility},} \textit{Econometric Reviews}, 27(1-3),
  46--78.

\bibitem[{Davis et~al.(2016)Davis, Zang, and Zheng}]{Davis16}
Davis, R.; Zang, P. and Zheng, T. (2016), \enquote{Sparse vector autoregressive
  modeling,} \textit{Journal of Computational and Graphical Statistics}, 25(4),
  1077--1096.

\bibitem[{Derimer et~al.(2017)Derimer, Diebold, Liu, and Yilmaz}]{Demirer17}
Derimer, M.; Diebold, F.~X.; Liu, L. and Yilmaz, K. (2017),
  \enquote{{Estimating Global Bank Network Connectedness},} \textit{Forthcoming
  in Journal of Applied Econometrics}.

\bibitem[{Diebold and Mariano(1995)}]{Diebold95}
Diebold, F.~X. and Mariano, R.~S. (1995), \enquote{{Comparing predictive
  accuracy},} \textit{Journal of Business and Economic Statistics}, 13(3),
  253--263.

\bibitem[{Diebold and Yilmaz(2014)}]{Diebold14}
Diebold, F.~X. and Yilmaz, K. (2014), \enquote{{On the network topology of
  variance decompositions: Measuring the connectedness of financial firms},}
  \textit{Journal of Econometrics}, 182(1), 119--134.

\bibitem[{Diebold and Yilmaz(2015)}]{Diebold15}
--- (2015), \textit{Financial and macroeconomics connectedness: A network
  approach to measurement and monitoring}, Oxford University Press, New York,
  US.

\bibitem[{Ding(2016)}]{Ding16}
Ding, P. (2016), \enquote{{On the conditional distribution of the multivariate
  t distribution},} \textit{The American Statistician}, 70(3), 293--295.

\bibitem[{Engle et~al.(2015)Engle, Jondeua, and Rockinger}]{Engle15}
Engle, R.; Jondeua, E. and Rockinger, M. (2015), \enquote{{Systemic risk in
  Europe},} \textit{Review of Finance}, 19 (1), 145--190.

\bibitem[{Finegold and Drton(2011)}]{Finegold11}
Finegold, M. and Drton, M. (2011), \enquote{Robust graphical modeling of gene
  networks using classical and alternative t-distributions,} \textit{The Annals
  of Applied Statistics}, 5(2A), 1057--1080.

\bibitem[{Franses and Lucas(1998)}]{Franses98}
Franses, P. and Lucas, A. (1998), \enquote{{Outlier detection in cointegration
  analysis},} \textit{Journal of Business \& Economic Statistics}, 16(4),
  459--468.

\bibitem[{Friedman et~al.(2007)Friedman, Hastie, H{\"o}fling, and
  Tibshirani}]{Friedman07}
Friedman, J.; Hastie, T.; H{\"o}fling, H. and Tibshirani, R. (2007),
  \enquote{Pathwise Coordinate Optimization,} \textit{The Annals of Applied
  Statistics}, 1(2), 302--332.

\bibitem[{Friedman et~al.(2008)Friedman, Hastie, and Tibshirani}]{Friedman08}
Friedman, J.; Hastie, T. and Tibshirani, R. (2008), \enquote{Sparse inverse
  covariance estimation with the graphical lasso,} \textit{Biostatistics},
  9(3), 432--441.

\bibitem[{Gelper et~al.(2016)Gelper, Wilms, and Croux}]{Gelper16}
Gelper, S.; Wilms, I. and Croux, C. (2016), \enquote{Identifying demand effects
  in a large network of product categories,} \textit{Journal of Retailing},
  92(1), 25--39.

\bibitem[{Hafner and Herwartz(2006)}]{Hafner06}
Hafner, C.~M. and Herwartz, H. (2006), \enquote{{Volatility impulse responses
  for multivariate GARCH models: An exchange rate illustration},}
  \textit{Journal of International Money and Finance}, 25(5), 719--740.

\bibitem[{Hassler et~al.(2016)Hassler, Rodrigues, and Rubia}]{Hassler16}
Hassler, U.; Rodrigues, P. M.~M. and Rubia, A. (2016), \enquote{{Quantile
  regression for long memory testing: A case of realized volatility},}
  \textit{Journal of Financial Econometrics}, 14(4), 693--724.

\bibitem[{Hastie et~al.(2015)Hastie, Tibshirani, and Wainwright}]{Hastie15}
Hastie, T.; Tibshirani, R. and Wainwright, M. (2015), \textit{Statistical
  learning with sparsity: The lasso and generalizations}, CRC press.

\bibitem[{Hautsch et~al.(2015)Hautsch, Schaumburg, and Schienle}]{Hautsch15}
Hautsch, N.; Schaumburg, J. and Schienle, M. (2015), \enquote{{Financial
  network systemic risk contributions},} \textit{Review of Finance}, 19(2),
  685--738.

\bibitem[{Knittel and Pindyck(2016)}]{Knittel16}
Knittel, C.~R. and Pindyck, R.~S. (2016), \enquote{{The simple economics of
  commodity price speculation},} \textit{American Economic Journal:
  Macroeconomics}, 8(2), 85--110.

\bibitem[{Kotz and Nadarajah(2004)}]{Kotz04}
Kotz, S. and Nadarajah, S. (2004), \textit{Multivariate \textit{t}
  distributions and their applications}, Cambridge University Press, Cambridge,
  UK.

\bibitem[{Lanne and Nyberg(2016)}]{Lanne16}
Lanne, M. and Nyberg, H. (2016), \enquote{{Generalized forecast error variance
  decomposition for linear and nonlinear multivariate models},} \textit{Oxford
  Bulletin of Economics and Statistics}, 78(4), 595--603.

\bibitem[{Levin et~al.(2002)Levin, Lin, and Chu}]{Levin02}
Levin, A.; Lin, C.-F. and Chu, C.-S.~J. (2002), \enquote{Unit root tests in
  panel data: Asymptotic and finite-sample properties,} \textit{Journal of
  Econometrics}, 108(1), 1--24.

\bibitem[{Liu and Rubin(1995)}]{Liu95}
Liu, C. and Rubin, D.~B. (1995), \enquote{{ML} estimation of the \textit{t}
  distribution using {EM} and its extensions, {ECM} and {ECME},}
  \textit{Statistica Sinica}, 5, 19--39.

\bibitem[{L\"{u}tkepohl(2005)}]{Lutkepohl05}
L\"{u}tkepohl, H. (2005), \textit{New introduction to multiple time series
  analysis}, Springer, Heidelberg, Germany.

\bibitem[{Martens and van Dick(2007)}]{Martens07}
Martens, M. and van Dick, D. (2007), \enquote{{Measuring volatility with the
  realized range},} \textit{Journal of Econometrics}, 138(1), 181--207.

\bibitem[{McAleer and Medeiros(2008)}]{McAleer08}
McAleer, M. and Medeiros, M.~C. (2008), \enquote{{Realized volatility: A
  review},} \textit{Econometric Reviews}, 27(1-3), 10--45.

\bibitem[{Parkinson(1980)}]{Parkinson80}
Parkinson, M. (1980), \enquote{{The extreme value method for estimating the
  variance of the rate of return},} \textit{The Journal of Business}, 53(1),
  61--65.

\bibitem[{Pesaran and Shin(1998)}]{Pesaran98}
Pesaran, H.~H. and Shin, Y. (1998), \enquote{{Generalized impulse response
  analysis in linear multivariate models},} \textit{Economics Letters}, 58(1),
  17--29.

\bibitem[{Rezitis(2015)}]{Rezitis15}
Rezitis, A.~N. (2015), \enquote{The relationship between agricultural commodity
  prices, crude oil prices and {US} dollar exchange rates: A panel {VAR}
  approach and causality analysis,} \textit{International Review of Applied
  Economics}, 29(3), 403--434.

\bibitem[{Rothman et~al.(2010)Rothman, Levina, and Zhu}]{Rothman10}
Rothman, A.~J.; Levina, E. and Zhu, J. (2010), \enquote{Sparse multivariate
  regression with covariance estimation,} \textit{Journal of Computational and
  Graphical Statistics}, 19 (4), 947--962.

\bibitem[{Serra(2011)}]{Serra11}
Serra, T. (2011), \enquote{Volatility spillovers between food and energy
  markets: A semiparametric approach,} \textit{Energy Economics}, 33,
  1155--1164.

\bibitem[{Serra and Zilberman(2013)}]{Serra13}
Serra, T. and Zilberman, D. (2013), \enquote{{Biofuel-related price
  transmission literature: A review},} \textit{Energy Economics}, 37, 141--151.

\bibitem[{Shu and Zhang(2006)}]{Shu06}
Shu, J. and Zhang, J.~E. (2006), \enquote{{Testing range estimators of
  historical volatility},} \textit{The Journal of Futures Markets}, 26(3),
  297--313.

\bibitem[{Tibshirani(1996)}]{Tibshirani96}
Tibshirani, R. (1996), \enquote{Regression shrinkage and selection via the
  lasso,} \textit{Journal of the Royal Statistical Society: Series B
  (Statistical Methodology)}, 58(1), 267--288.

\end{thebibliography}

\end{document}